\documentclass[%
reprint,
superscriptaddress,
nofootinbib,
 amsmath,amssymb,
 aps,
]{revtex4-2}

\usepackage{graphicx}
\usepackage{dcolumn}
\usepackage{bm}
\usepackage{hyperref}
\usepackage[mathlines]{lineno}
\usepackage{amsmath,amssymb,euscript}
\usepackage{calrsfs}

\usepackage{tikz}
\usetikzlibrary{calc,decorations.markings}
\usetikzlibrary{snakes}
\usepackage{rotating}
\usepackage{comment}
\usepackage{empheq}

\allowdisplaybreaks[2]
\numberwithin{equation}{section}

\definecolor{darkgreen}{rgb}{0.0, 0.55, 0.1}

\begin{document}

\title{Tidal deformations of a binary system induced by an external Kerr black hole}

\author{Filippo Camilloni}
\affiliation{Dipartimento di Fisica e Geologia, Universit\`a di Perugia, I.N.F.N. Sezione di Perugia, \\ Via Pascoli, I-06123 Perugia, Italy}
\author{Gianluca Grignani}
\affiliation{Dipartimento di Fisica e Geologia, Universit\`a di Perugia, I.N.F.N. Sezione di Perugia, \\ Via Pascoli, I-06123 Perugia, Italy}
\author{Troels Harmark}
 \affiliation{Niels Bohr International Academy, Niels Bohr Institute, Copenhagen University,\\ Blegdamsvej 17, DK-2100 Copenhagen \O{}, Denmark}
\author{Roberto Oliveri}%
\affiliation{LUTH, Laboratoire Univers et Th\'eories, Observatoire de Paris,\\ CNRS, Universit\'e PSL, Universit\'e Paris Cit\'e,\\ 5 place Jules Janssen, 92190 Meudon, France}
\author{Marta Orselli}
 \affiliation{Dipartimento di Fisica e Geologia, Universit\`a di Perugia, I.N.F.N. Sezione di Perugia, \\ Via Pascoli, I-06123 Perugia, Italy}
 \affiliation{Niels Bohr International Academy, Niels Bohr Institute, Copenhagen University,\\ Blegdamsvej 17, DK-2100 Copenhagen \O{}, Denmark}
 \author{Daniele Pica}
 \affiliation{Dipartimento di Fisica e Geologia, Universit\`a di Perugia, I.N.F.N. Sezione di Perugia, \\ Via Pascoli, I-06123 Perugia, Italy}
 \affiliation{Niels Bohr International Academy, Niels Bohr Institute, Copenhagen University,\\ Blegdamsvej 17, DK-2100 Copenhagen \O{}, Denmark}

\begin{abstract}
The dynamics of a binary system moving in the background of a black hole is affected by tidal forces.
In this work, for the Kerr black hole, we derive the electric and magnetic tidal moments at quadrupole order, where the latter are computed for the first time in full generality. 
We make use of these moments in the scenario of a hierarchical triple system made of a Kerr black hole and an extreme-mass ratio binary system consisting of a Schwarzschild black hole and a test particle. We study how the secular dynamics of the test particle in the binary system is distorted by the presence of tidal forces from a much larger Kerr black hole.
Our treatment includes strong gravitational effects beyond the post-Newtonian approximation both for the binary system and for the tidal forces since the binary system is allowed to be close to the event horizon of the Kerr black hole. 
We compute the shifts in the physical quantities for the secular dynamics of the test particle and show that they are gauge invariant. In particular, we apply our formalism to the innermost stable circular orbit for the test particle and to the case of the photon sphere.
Our results are relevant for the astrophysical situation in which the binary system is in the vicinity of a supermassive black hole.
\end{abstract}

\maketitle
\tableofcontents

\section{Introduction}  
The detection of gravitational waves from coalescing binary systems by the LIGO-Virgo-Kagra collaboration \cite{LIGOScientific:2016aoc, LIGOScientific:2020ufj,LIGOScientific:2021qlt} has unsealed a new powerful and fascinating way of exploring our Universe in a regime of strong gravitational field.  
This has made it increasingly relevant to investigate new types of strong gravitational phenomena analytically, to prepare for future experimental results.

Indeed, with the next generation detectors such as the ground-based Einstein Telescope~\cite{Maggiore:2019uih} and Cosmic Explorer~\cite{Evans:2021gyd}, as well as
the space-based  LISA~\cite{LISA:2017pwj} and TianQin~\cite{TianQin:2020hid}, 
the sensitivity and frequency band  will be greatly expanded.
This will make it possible  to use black hole binary systems also as probes of their surrounding environment (see Ref.~\cite{Barausse:2014tra} for a comprehensive review).
Examples of the effect of the environment include the presence of various types of energy and matter, such as an accretion disc \cite{Kocsis:2011dr, Derdzinski:2020wlw, Speri:2022upm} or dark matter \cite{Gondolo:1999ef, Bertone:2004pz, Macedo:2013qea, Eda:2014kra, Hannuksela:2019vip, Coogan:2021uqv, DeLuca:2021ite,DeLuca:2022xlz,Cole:2022fir,Kim:2022mdj}. 
Another example, relevant for this paper, is the presence of a third body, such as a nearby supermassive black hole \cite{Bonetti:2016eif,Bonetti:2017dan,Bonetti:2017lnj,Bonetti:2018tpf,PhysRevD.83.044030,PhysRevD.102.104002,Yang:2017aht,Bonga:2019ycj,Peng:2021vzr,Chen:2022tdj} bound to the binary system.

Moreover, the expansion in sensitivity and frequency band will make it possible to detect signals from new types of sources, such as for example extreme-mass-ratio (EMR) inspiraling systems.~\footnote{We adopt hereafter the abbrevation EMR instead of the more conventional EMRI to stress that, in this work, we do not focus on inspiral phases and we neglect any radiation-reaction effects in comparison to tidal effects.} Among these systems, the ones that will typically be detectable in the LISA band \cite{Barausse:2020rsu, Amaro-Seoane:2010dmp}, are made of a stellar mass compact object of mass $m$ and a black hole with a much larger mass $M\gg m$, with mass ratios $m/M$ ranging from $10^{-4}$ to $10^{-6}$. 

In this paper we are interested in the dynamical effects of having a binary black hole system immersed in a curved background spacetime. 
To access a scenario that is at the same time realistic, includes strong gravitational effects, and can be treated analytically, we consider the case of an EMR binary system, \emph {i.e.} a black hole and a test particle, in the background of a third, larger black hole, affecting the binary system through tidal forces. 

We take the curved background spacetime to be the general case of a Kerr black hole of mass $M_*$. Instead the EMR binary system will consist of a Schwarzschild black hole of mass $M$ with a test particle of mass $m$, enabling us to use the 
tidally deformed Schwarzschild metric of Refs.~\cite{PhysRevLett.94.161103,Poisson:2009qj} to describe the EMR binary system. 
For the test particle we consider it to move on a geodesic, neglecting higher order effects in $m/M$ such as the self-force. 
As the size of the binary system will be set by the scale $M$, we need $M \ll \mathcal{R}$ where $\mathcal{R}$ is the curvature length scale set by the background Kerr black hole. This ensures that the effects of the background can be described through tidal forces, with the condition $M \ll \mathcal{R}$ known as the {\it small-tide approximation}~\cite{Poisson:2009qj}. 

We will consider the quadrupole approximation to the tidal forces, being the leading order in $M/\mathcal{R}$. 
This means we can consider the EMR binary as moving on a geodesic of the Kerr black hole geometry.  
A particularly interesting regime is when $M_* \gg M$ thus corresponds to a hierarchical three body system. In this case, the binary system can be close to the event horizon of the Kerr black hole, even while the small-tide approximation is respected. 

Our setup is inspired by that of Ref.~\cite{Yang:2017aht}, while at the same time being a significant extension. Their setup was restricted to a Schwarzschild black hole as the third body, and the EMR binary system was assumed to be at a large distance. Instead, we are able to consider the strong gravitational effects on the binary system when it moves in close vicinity to a Kerr black hole. This also means that we need to consider more carefully the relative orientation of the EMR binary system relative to the Kerr black hole. This is accomplished by introducing two independent rotation angles. 
Moreover, it is important to note that in our setup we are able to capture strong 
gravitational effects arising from curved spacetime, in contrast with most of the extensive literature on three body systems \cite{1987PhLA..123..336S,Konigsdorffer:2003ue,Lousto:2007ji,Torigoe:2009bw,Galaviz:2010te}, as those works employ the approximation that all three bodies are small relative to their mutual distances. 

A significant part of our paper concerns the careful computation of the general quadrupole tidal forces due to the Kerr black hole, as these constitute the forces that can affect the binary system in our setup. These forces are described by the tidal tensors $C_{ij}$ and $C_{ijk}$. The rank-2 tidal tensors $C_{ij}$  were previously computed for a generic value of the Kerr angle $\hat{\theta}$ in a seminal paper by Marck~\cite{Marck:1973}, where he constructed the orthonormal tetrad that is parallel-transported  along an arbitrary time-like geodesic in the Kerr spacetime. From the rank-2 tidal tensors $C_{ij}$ one can then compute the ``electric'' quadrupole moments $\mathcal{E}_{ij}$, which can be considered as ``mass moments" produced by gravitational forces external to a certain region.

A primary result of this paper, is the derivation of the general form of the rank-3 tidal tensors $C_{ijk}$
for all values of the angle $\hat{\theta}$ in the Kerr spacetime. This generalizes the results of Ref.~\cite{Alvi:1999cw} (later confirmed in Ref.~\cite{Poisson:2003nc}), where the tidal tensors $C_{ijk}$ were obtained only for the specific value $\hat{\theta}=\pi/2$, namely in the equatorial plane of the Kerr spacetime.
From the rank-3 tidal tensors $C_{ijk}$ we moreover derive the ``magnetic'' quadrupole moments $\mathcal{B}_{ij}$, which can be considered as external ``current moments" and generate velocity-dependent tidal forces on test bodies. This is another original result of this paper.

We apply these tidal electric and magnetic quadrupole moments to the case described above, with an EMR binary system following a geodesic in the Kerr background. 
The effects induced by the tidal fields can be studied by computing the Hamiltonian of a test particle (the object of mass $m$) in the tidally deformed Schwarzschild  spacetime. 
Specifically, starting from a circular orbit in the unperturbed Schwarzschild spacetime, we find that the geodesics in the tidally deformed spacetime acquire a small  eccentricity proportional to the deformation parameter. The quasi-circular dynamics in the perturbed spacetime is governed by a {\it secular Hamiltonian}, which keeps into account the effects of the tidal deformation on circular orbits. It can be written as a sum of the unperturbed Hamiltonian in the Schwarzschild spacetime and an interaction term of order $\sim M/M_*$, which allows us, for example, to compute perturbatively the effects of tides on the location and properties of the innermost stable circular orbit (ISCO) and of the photon sphere. 

Using the tidal moments we computed, we derive the effects of tides on the frequency, radius, energy and angular momentum  of the ISCO of the binary system, by computing the shifts induced by the small tides on these physical quantities.~\footnote{See Ref.~\cite{Isoyama:2014mja} for similar treatments in the context of the self-force approximation.} The case of tides generated by a Schwarzschild black hole was studied in Refs.~\cite{Yang:2017aht,Cardoso:2021qqu}. Here we derive the shifts in the case of tides induced by the Kerr geometry and we derive the expression of the parameter $\eta$ entering these shifts. We find that $\eta$ depends on the spin of the Kerr black hole, the Carter constant $K$, the Kerr angle $\hat{\theta}$ and the Boyer-Lindquist radius $\hat{r}$ at which the black hole of mass $M$ is located in the Kerr spacetime geometry.
More generally, our result does not rely on the specific nature of the third body responsible for the tides. Indeed, the tidal parameter $\eta$ in the secular Hamiltonian is shown to be proportional to the secular average of the scalar part of the electric tidal moment. This result holds in the quadrupole and in the secular approximation. We provide an expression for $\eta$ in terms of arbitrary tides and specialize it to the case of a Kerr black hole.

The paper is organized as follows. In Sec~\ref{sec: tidal construction}, we compute the tidal moments induced by a Kerr black hole. Following Ref.~\cite{Marck:1973}, we first recover the already known expression for the electric tidal moments and then we derive the most general expressions for the magnetic components of the tidal moments, generalising the computation done in Ref.~\cite{Alvi:1999cw}. In Sec.~\ref{Sec:triplete}, we introduce the hierarchical triple system that we analyse in this paper. We write down the metric for a tidally deformed Schwarzschild black hole up to the quadrupole order. 
We moreover write down the explicit expression for the quadrupole electric and magnetic moments and we introduce the Euler angles which allow us to study any possible orientation of the binary system. In Sec.~\ref{sec: secular H}, we focus on the secular dynamics of the binary system in order to understand how the parameters which specify the orbits of the test particle around the Schwarzschild black hole, such as energy and angular momentum, are shifted by the tidal fields. In Sec.~\ref{sec: secular shift}, we apply the results of the previous sections to the case in which the test particle is moving along the ISCO of the Schwarzschild black hole. 
In addition, we extend our computation also to the case of a massless particle studying how the photon sphere is deformed by the tidal fields. We furthermore discuss the gauge invariance of our results.
Finally, Sec~\ref{Concl} contains our concluding remarks.

\bigskip
Throughout this paper Greek indices run from 0 to 3, Latin lower-case indices ($i,j,k,...$) run from 1 to 3, Latin upper-case indices ($A, B, C, ...$) label spherical coordinates. Indices in round brackets ($(a), (b), (c), ...$) label tensor components in the Carter's tetrad.
Symmetric and tracefree (STF) tensors are denoted by angular brackets over their indices, \emph{e.g.}, $T_{\langle i j\rangle} = T_{(ij)} - \frac{1}{3}\delta_{ij}T_{kl}\delta^{kl}$.
Hatted coordinates $(\hat t,\hat r, \hat \theta, \hat \phi)$ are employed for the Kerr spacetime. Schwarzschild coordinates, used for the binary system, are instead denoted as $(t,r,\theta,\phi)$.
We use geometrized units with $G=c=1$ and the Minkowski metric signature is $\eta=\text{diag}(-1,1,1,1)$. 

\section{Tidal moments induced by a Kerr black hole}
\label{sec: tidal construction}
In this section we derive the general quadrupole tidal moments for geodesic motion around a Kerr black hole which we will use in Sections~\ref{Sec:triplete}-\ref{sec: secular shift}. In Sec.~\ref{sec: CarterTetrad} we define the \emph{Carter's tetrad}, in terms of which the curvature tensor simplifies. In Sec.~\ref{sec: MarckTetrad} we present an alternative inertial frame \cite{Marck:1973}, parallel-transported along a generic geodesic in the Kerr spacetime, here called the \emph{Marck's tetrad}. 
This is the most suitable reference frame in which it is possible to extract analytic information concerning the tidal effects induced by the Kerr geometry on a system moving along its geodesics. The tidal effects are encoded in the rank-2 and rank-3 tidal tensors and in the set of electric and magnetic tidal  moments, explicitly given in Sec.~\ref{sec: tidaltensors} and \ref{sec: tidalmoments} at the quadrupole order. The expressions of the  rank-3 tidal tensor and of the magnetic quadrupole moments outside the Kerr equatorial plane are derived here for the first time.

\subsection{Carter's tetrad}
\label{sec: CarterTetrad}
The Kerr metric for a rotating black hole of mass $M_*$ and spin $J_*$,  in Boyer-Lindquist (BL) coordinates $\hat x^\mu=(\hat t,\hat r,\hat \theta,\hat \phi)$ takes the form 
\begin{equation} 
    \label{KerrBL}
    \begin{split}
        d\hat s^2=-\Big(1-&\frac{2M_*\hat r}{\Sigma}\Big)d\hat t^2-\frac{4M_*\hat r}{\Sigma}a\sin^2 \hat\theta ~d\hat t\, d\hat \phi
        \\
        &+\frac{\mathcal{A}}{\Sigma}\sin^2 \hat \theta ~d\hat \phi^2
        +\frac{\Sigma}{\Delta}d\hat r^2+\Sigma d\hat \theta^2~,
    \end{split}
\end{equation}
where $a=J_*/M_*$ is the specific angular momentum  and 
\begin{equation}
\label{sigma}
    \begin{split}
        \Sigma = \hat r^2&+a^2 \cos^2 \hat \theta, \quad \Delta=\hat r^2- 2M_* \hat r+a^2 ,
        \\
        &\mathcal{A} = (\hat r^2+a^2)^2-a^2\Delta\sin^2 \hat \theta~.
    \end{split}
\end{equation}
We are interested in considering time-like geodesics around a Kerr black hole, specified by three constants of motion: the energy per unit mass $\hat E$, the angular momentum per unit mass $\hat L$ and the Carter constant $K$.
More specifically, the first integrals of the equations of motion read \cite{PhysRev.174.1559}
\begin{align}\label{eq: Kerr_geod}
\begin{split}
    \dot{\hat t} &=\frac{\mathcal{A} \hat E - 2M_* \hat r a \hat L }{\Delta \Sigma}~,
    \\
    \dot{\hat r}^2 &= \left[ \frac{\hat E ({\hat r}^2+ a^2) - a \hat L}{\Sigma}\right]^2 - \frac{\Delta}{\Sigma^2} (\hat r^2+K)~,
    \\
     \dot{\hat \theta}^2 &= \frac{1}{\Sigma^2} \left[K - a^2 \cos \hat \theta - \bigg(a \hat E \sin \hat \theta - \frac{\hat L}{\sin \hat \theta} \bigg)^2\right]~,
    \\
    \dot{\hat \phi} &= \frac{1}{\Delta}\left[\frac{2 M_* \hat r a\hat E }{\Sigma} + \left( 1 - \frac{2M_*\hat r}{\Sigma }\right) \frac{\hat L}{ \sin^2\hat \theta}\right]~,
\end{split}
\end{align}
where the dot denotes differentiation with respect to the proper time $\tau$.

A convenient tetrad for the Kerr geometry \eqref{KerrBL}, such that $d\hat{s}^2 = \eta_{(a)(b)} \omega^{(a)}\omega^{(b)}$, was introduced in Ref.~\cite{cmp/1103841118} and reads
\begin{align} 
\label{eq:Cartertetrad}
    \begin{split}
        \omega^{(0)}&=\sqrt{\frac{\Delta}{\Sigma}}\left(d\hat{t}-a \sin^2\hat{\theta} d\hat{\phi}\right)~,
        \\
        \omega^{(1)}&=\sqrt{\frac{\Sigma}{\Delta}}d\hat{r}~,
        \\ 
        \omega^{(2)}&=\sqrt{\Sigma}d\hat{\theta}~,
        \\
        \omega^{(3)}&=\frac{\sin\hat{\theta}}{\sqrt{\Sigma}}\left(a d\hat{t}-(\hat{r}^2+a^2) d\hat{\phi}\right)~.
    \end{split}
\end{align}

We dub this tetrad, the {\it{Carter's tetrad}}. The curvature 2-form
\begin{equation}
    \Omega_{(a)(b)} = \frac{1}{2}C_{(a)(b)(c)(d)}\omega^{(c)}\wedge \omega^{(d)}\,,
\end{equation}
with $C_{(a)(b)(c)(d)}$ being the components of the Weyl tensor, [$C_{\mu\nu\rho\sigma}=R_{\mu\nu\rho\sigma}$ for the Kerr geometry \eqref{KerrBL}], projected along the Carter's tetrad with the inverses of Eq.~\eqref{eq:Cartertetrad}, $\omega^\mu_{(a)}$,  $C_{(a)(b)(c)(d)}=C_{\mu\nu\rho\sigma}~\omega^\mu_{(a)}\omega^\nu_{(b)}\omega^\rho_{(c)}\omega^\sigma_{(d)}$, reads~
\cite{Carter:1973rla,Marck:1973}
\begin{align}
\begin{split}
        \Omega^{(0)(1)} &= 2I_1~ \omega^{(0)}\wedge \omega^{(1)} + 2I_2~ \omega^{(2)}\wedge \omega^{(3)}~,\\
        \Omega^{(0)(2)} &= -I_1~ \omega^{(0)}\wedge \omega^{(2)} + I_2~ \omega^{(1)}\wedge \omega^{(3)}~,\\
        \Omega^{(0)(3)} &= -I_1~ \omega^{(0)}\wedge \omega^{(3)} - I_2~ \omega^{(1)}\wedge \omega^{(2)}~,\\
        \Omega^{(1)(2)} &= -I_1~ \omega^{(1)}\wedge \omega^{(2)} + I_2~ \omega^{(0)}\wedge \omega^{(3)}~,\\
        \Omega^{(1)(3)} &= -I_1~ \omega^{(1)}\wedge \omega^{(3)} - I_2~ \omega^{(0)}\wedge \omega^{(2)}~,\\
        \Omega^{(2)(3)} &= 2I_1~ \omega^{(2)}\wedge \omega^{(3)} - 2I_2~ \omega^{(0)}\wedge \omega^{(1)}~,
\end{split}
\end{align}
where
\begin{equation}
\label{I1I2}
    \begin{split}
        I_1&= \frac{M_* \hat{r}}{\Sigma^3} \left(\hat{r}^2-3 a^2 \cos ^2\hat{\theta} \right) ~,
        \\
        I_2&=
        \frac{a M_* \cos \hat{\theta} }{\Sigma^3} \left(3 \hat{r}^2-a^2 \cos ^2\hat{\theta} \right)~.
    \end{split}
\end{equation}
%

\subsection{Marck's tetrad}
\label{sec: MarckTetrad}
The orthonormal tetrad $\lambda^{(a)} = \left(\lambda_0^{(a)}, \lambda_1^{(a)}, \lambda_2^{(a)}, \lambda_3^{(a)}\right)$ that is parallel-transported  along an arbitrary time-like geodesic was constructed in Ref.~\cite{Marck:1973}. The tetrad component $\lambda_0^{(a)}$ is a time-like unit vector tangent to the geodesics and $\lambda_i^{(a)}$ are space-like unit vectors. They satisfy the following conditions 
\begin{equation}\label{Marckcond}
    \eta_{(a)(b)}~\lambda_\alpha^{(a)} \lambda_\beta^{(b)}=\eta_{\alpha\beta}~,\quad \lambda_0^{\mu}\nabla_{\mu}\lambda_\alpha^{\nu} = 0~,
\end{equation}
where $\lambda^\mu_\alpha=\omega^\mu_{(a)}\lambda^{(a)}_\alpha$ and $\alpha,\beta = \{0,1,2,3\}$ are the labels of the components of the tetrad.
The first relation in Eq.~\eqref{Marckcond} is the orthonormal condition, the second one is the parallel-transported requirement that implies the tetrad frame is inertial. 
Their explicit expressions in terms of the metric functions and the constants of motion are~\cite{Marck:1973}~\footnote{We rename $\lambda_{2}^{(a)}$ and $\tilde{\lambda}_{3}^{(a)}$ in Ref.~\cite{Marck:1973} with our $\lambda_{3}^{(a)}$ and $\tilde{\lambda}_{2}^{(a)}$, respectively. It is also important to stress that all the components of the space-like vectors $\lambda_i^{(a)}$ can be written in terms of $\lambda_0^{(a)}$.}
\begin{align}
\label{MarckTetrad}
\begin{split}
    \lambda_0^{(a)} &= \left(\frac{\hat E (\hat r ^2+a^2)-a\hat L}{\sqrt{\Delta\Sigma}}, \sqrt{\frac{\Sigma}{\Delta}}\dot{\hat r}, \sqrt{\Sigma}\dot{\hat \theta}, \frac{a\hat E\sin^2 \hat \theta-\hat L}{\sin\hat\theta\sqrt{\Sigma}}\right)~,\\
    \lambda_1^{(a)}&=\tilde\lambda_1^{(a)}\cos\Psi-\tilde\lambda_2^{(a)}\sin\Psi~,\\
    \lambda_2^{(a)}&=\tilde\lambda_1^{(a)}\sin\Psi+\tilde\lambda_2^{(a)}\cos\Psi~,\\
    \lambda_3^{(a)}&= \frac{1}{\sqrt{K}} \left(a \cos\hat{\theta}\lambda_0^{(1)},
    a \cos\hat{\theta}\lambda_0^{(0)},
    - \hat{r}\lambda_0^{(3)},
    \hat{r}\lambda_0^{(2)}\right)~,
\end{split}
\end{align}
where
\begin{align}
\begin{split}
    \tilde\lambda_1^{(a)}&=\sqrt{\frac{T}{KS}}\left(\hat{r}\lambda_0^{(1)},
    \hat{r}\lambda_0^{(0)},
    \frac{S}{T}a\cos{\hat{\theta}}\lambda_0^{(3)},
    -\frac{S}{T} a\cos{\hat{\theta}}\lambda_0^{(2)}\right),\\
    \tilde\lambda_2^{(a)}&=\sqrt{\frac{T}{S}}\left(\lambda_0^{(0)},\lambda_0^{(1)},\frac{S}{T}\lambda_0^{(2)},\frac{S}{T}\lambda_0^{(3)}\right)~,
\end{split}
\end{align}
and
\begin{equation}\label{ST}
    S=\hat{r}^2+K~,\quad T=K-a^2\cos^2\hat{\theta}~.
\end{equation}
Notice the identity $\Sigma = S-T$. 
In the second and third tetrad component of Eq.~\eqref{MarckTetrad}, we rotated the vectors $\tilde{\lambda}_1^{(a)}$ and $\tilde{\lambda}_2^{(a)}$ of an angle $\Psi$. This is necessary in order to ensure that the tetrad $\lambda^{(a)} = \left(\lambda_0^{(a)}, \lambda_1^{(a)}, \lambda_2^{(a)}, \lambda_3^{(a)}\right)$ is parallel-transported along the geodesic motion \cite{Marck:1973}. 
In particular $\Psi$ is an angle depending on the proper time along the Kerr geodesic. The equation satisfied by $\Psi$ was derived in Ref.~\cite{Marck:1973} and reads
\begin{equation}
\label{general_psi_dot}
    \dot\Psi=\frac{\sqrt{K}}{\Sigma}\left(\frac{\hat{E}(\hat{r}^2+a^2)-a \hat{L}}{S}+a \frac{\hat{L}-a \hat{E}\sin^2\hat{\theta}}{T}\right)~.
\end{equation}
A solution for this first order differential equation was provided in Ref.~\cite{Marck:1973} and, more explicitly in terms of the Mino time, in Ref.~\cite{vandeMeent:2019cam}.

\subsection{Tidal tensors}
\label{sec: tidaltensors}
Tidal effects on test particles moving in the neighborhood of a geodesic in Kerr spacetime are best computed by evaluating the Weyl tensor on the parallel-transported tetrad $\lambda^{(a)}$ (see Eq.~\eqref{MarckTetrad}) with $\lambda^{(a)}_0$ being the four-velocity.
The explicit expressions for the tidal tensors are obtained once the Weyl tensor $C_{\mu\nu\rho\sigma}$ is evaluated on the Kerr geodesic. In order to compute the electric and magnetic quadrupole moments, we first need the following components of the rank-$2$ and rank-$3$ tidal tensors in the basis of the tetrad $\lambda^{(a)}$~\cite{Marck:1973,Poisson:2009qj}
\begin{align}
\begin{split}
    C_{ij}&\equiv  C_{(a)(b)(c)(d)}\lambda^{(a)}_0\lambda^{(b)}_i\lambda^{(c)}_0\lambda^{(d)}_j~,
    \\
    C_{ijk}&\equiv C_{(a)(b)(c)(d)} \lambda^{(a)}_0\lambda^{(b)}_i\lambda^{(c)}_j\lambda^{(d)}_k ~,
\end{split}
\end{align}
where we recall that $C_{(a)(b)(c)(d)} = C_{\mu\nu\rho\sigma}~{\omega^\mu}_{(a)}{\omega^\nu}_{(b)}{\omega^\rho}_{(c)}{\omega^\sigma}_{(d)}$.
Note that, as a consequence of the symmetries of the Weyl tensor, $C_{ij}$ is an STF tensor, whereas $C_{ijk}$ is trace-free and anti-symmetric in $(j,k)$ by definition. Moreover, it obeys the condition $C_{ijk} + C_{jki} + C_{kij} =0$, implying that $C_{ijk} - C_{jik} = -C_{kij} $ and $C_{ijk} - C_{kji} = - C_{jki}$.

\begin{widetext}
We compute now the explicit expression for the components of the Weyl tensor that are relevant for the calculations of the electric and magnetic quadrupole moments. Our expressions are valid for arbitrary time-like geodesics in the Kerr black hole spacetime. The $C_{ij}$  read 
\begin{align}\label{Cij_Kerr}
\begin{split}
C_{11}&=\left[1-\frac{3ST}{K \Sigma^2}(\hat{r}^2-a^2 \cos^2\hat{\theta})\cos^2\Psi\right]I_1
+\frac{6ST}{K \Sigma^2} a \hat{r}\cos\hat{\theta}\cos^2\Psi I_2~,
\\
C_{12}&=-\frac{3 ST}{K\Sigma^2}\left[\left(\hat{r}^2-a^2\cos^2\hat{\theta}\right)I_1
-2 a \hat{r}\cos\hat{\theta} I_2\right]\sin\Psi\cos\Psi~,\\
C_{13}&= - \frac{3 \sqrt{S T}  }{K \Sigma ^2}\left[ a\hat{r}\cos\hat{\theta}(S+T)I_1
+ \left(\hat{r}^2 T - a^2 S\cos^2\hat{\theta}\right) I_2\right]\cos\Psi~,\\
C_{22}&=\left[1-\frac{3ST}{K \Sigma^2}(\hat{r}^2-a^2 \cos^2\theta)\sin^2\Psi\right]I_1
+\frac{6ST}{K \Sigma^2} a \hat{r}\cos\hat{\theta}\sin^2\Psi I_2~,\\
C_{23}&=- \frac{3 \sqrt{S T}  }{K \Sigma ^2}\left[ a\hat{r}\cos\hat{\theta}(S+T)I_1
+ \left(\hat{r}^2 T - a^2 S\cos^2\hat{\theta}\right) I_2\right]\sin\Psi~,\\
C_{33}&=\left[1+\frac{3}{K \Sigma^2}(\hat{r}^2 T^2-a^2 S^2 \cos^2\hat{\theta})\right]I_1-\frac{6 ST}{K \Sigma^2} a \hat{r}\cos\hat{\theta} I_2~.
\end{split}
\end{align}
Note that $C_{ij}$ was already computed in Ref.~\cite{Marck:1973} (with the label $2$ renamed with $3$ in this paper).

As a new result, we provide also the general expression for the non-vanishing components of the rank-3 tidal tensor $C_{ijk}$ that enter the calculation of the magnetic moments which will be done in the next subsection. The non-vanishing components are given by 
\begin{align}\label{Cijk_Kerr}
\begin{split}
C_{112} &= \frac{3 \sqrt{S T}}{K \Sigma ^2}\left[ \left(\hat{r}^2 T-a^2 S \cos^2\hat{\theta} \right) I_1 -  a \hat{r} \cos \hat{\theta} (S+T) I_2 \right] \cos \Psi~,\\
C_{113} &= \frac{3 ST}{K \Sigma ^2} \left[ 2 a \hat{r}\cos \hat{\theta}  I_1 + \left(\hat{r}^2-a^2 \cos ^2\hat{\theta}\right)I_2 \right]\sin \Psi  \cos \Psi ~,\\
C_{123} &=-\frac{6 S T}{K \Sigma ^2} a \hat{r}  \cos \hat{\theta}  \cos ^2\Psi  I_1+\frac{1}{K\Sigma ^2}\left[\left(\hat{r}^2 T + a^2 S \cos ^2\hat\theta \right)\left(S-T\right) - 3 S T \left(\hat{r}^2-a^2\cos^2\hat{\theta}\right) \cos ^2\Psi\right] I_2~,\\
C_{212} &= \frac{3 \sqrt{S T}}{K \Sigma ^2} \left[ \left(\hat{r}^2 T-a^2 S \cos^2\hat{\theta}\right) I_1 - a \hat{r} \cos \hat{\theta} (S+T) I_2\right]\sin \Psi ~,\\
C_{213}&=\frac{6 S T}{K \Sigma ^2} a \hat{r}  \cos \hat{\theta} \sin ^2\Psi I_1 +\frac{1}{K \Sigma ^2}\left[\hat{r}^2 T (2 S+T)-a^2  \cos ^2\hat{\theta} S(S+2T)-3 S T  \left(\hat{r}^2-a^2\cos^2\hat\theta\right)\cos ^2\Psi \right]I_2,\\
C_{312} &= \frac{6  S T}{K \Sigma ^2} a \hat{r}\cos\hat{\theta} I_1 +\frac{1}{K\Sigma ^2}\left[\hat{r}^2 T (S+2 T)-a^2 \cos ^2\hat{\theta} S(2S+T)\right]I_2~.
\end{split}
\end{align}
In addition, we observe that $C_{223} = -C_{113}$, $C_{312} = C_{213}-C_{123}$, $C_{313} = -C_{212}$, $C_{323} = C_{112}$.
If we specialize to equatorial ($\hat{\theta}=\pi/2$) geodesics in Kerr, the explicit expressions for the tidal tensors simplify considerably. We get, in agreement with Refs.~\cite{Marck:1973,Poisson:2003wz, Alvi:1999cw}, 
\begin{align}\label{eq: C_Kerr}
\begin{split}
    C_{11}&=\left[ 1 - 3 \left( 1 + \frac{ K}{\hat{r}^2} \right) \cos^2 \Psi \right] \frac{M_*}{\hat{r}^3}~,
    \\
    C_{22}&=\left[ 1 - 3 \left( 1 + \frac{ K}{\hat{r}^2} \right) \sin^2 \Psi \right] \frac{M_*}{\hat{r}^3}~,
    \\
    C_{12}&=-3\left( 1 + \frac{K}{\hat{r}^2} \right)  \frac{M_*}{\hat{r}^3} \cos \Psi \sin \Psi~,
    \\
    C_{33}&= \left(1+3\frac{ K}{\hat{r}^2}\right) \frac{M_*}{\hat{r}^3}~,
\end{split}
\end{align}
\end{widetext}
and, for the rank-$3$ tidal tensor (in agreement with Refs.~\cite{Alvi:1999cw} and~\cite{Poisson:2003nc}),
\begin{align}
\begin{split}
    C_{121}&=- \frac{3M_*\sqrt{K}}{\hat{r}^4}\sqrt{1+\frac{K}{\hat{r}^2}} \cos \Psi~,
    \\
    C_{221}&=- \frac{3M_*\sqrt{K}}{\hat{r}^4}\sqrt{1+\frac{K}{\hat{r}^2}} \sin \Psi~,
\end{split}
\end{align}
with $C_{121}=-C_{112}=C_{332}=-C_{323}$ and $C_{221}=-C_{212}=C_{313}=-C_{331}$.
\\
We also recall that 
for circular geodesics in the equatorial plane of the Kerr spacetime, the following expressions hold~\cite{Bardeen:1972fi}
\begin{equation}
\label{eq: EL_Kerr}
\begin{split}
    \hat E &=\frac{\hat{r}^{3/2}-2 M_* \hat{r}^{1/2} + \sigma a  M_*^{1/2}}{\hat{r}^{3/4} \sqrt{\hat{r}^{3/2}-3M_* \hat{r}^{1/2} + 2\sigma a  M_*^{1/2} }}~,
    \\
    \hat L &=\frac{\sigma  M_*^{1/2} \left(\hat{r}^2+  a^2  -2 \sigma a ~M_*^{1/2} \hat{r}^{1/2} \right)}{\hat{r}^{3/4} \sqrt{\hat{r}^{3/2}-3M_* \hat{r}^{1/2}  + 2\sigma a  M_*^{1/2} } }~,
    \\
    K &= \left(a  \hat E-\hat L \right)^2~,
    \\
    \dot{\Psi} &= \frac{\sqrt{K}}{\hat{r}^2+K} \left(\hat E - \frac{a}{a\hat E-\hat L}\right) = \sigma \sqrt{\frac{M_*}{\hat{r}^3}}~.
\end{split}
\end{equation}
Above we introduced the parameter $\sigma=\pm 1$ that allows one to distinguish between prograde ($+$) and retrograde ($-$) orbits.
A thorough analysis of the dynamics in the equatorial plane will be given in Sec.~\ref{Sec:equatorial}.


\subsection{Electric and magnetic quadrupole moments}
\label{sec: tidalmoments}
The \emph {electric} and \emph{magnetic quadrupole moments} in Cartesian coordinates are defined as \cite{Poisson:2009qj}
\begin{equation}\label{EB}
    \mathcal{E}_{ij}\equiv C_{ij}~,\quad \mathcal{B}_{ij} \equiv -\frac{1}{2}\epsilon_{k l\langle i} C^{~~kl}_{j\rangle}~,
\end{equation}
with $\epsilon_{ijk}$ the three-dimensional Levi-Civita symbol with $\epsilon_{123}=+1$. We raise and lower Cartesian indices ($i,j,k,...$) with the Kronecker delta $\delta_{ij}$. Being STF tensors, both the electric $\mathcal{E}_{ij}$ and the magnetic $\mathcal{B}_{ij}$ tensors have each five independent components thus, together, they account for the ten independent components of the Weyl tensor.
In particular, the magnetic quadrupole moments in terms of the components of the rank-3 tidal tensor, read
\begin{equation}
\begin{split}
    \mathcal{B}_{11}&=-C_{123}~,~~
    \mathcal{B}_{12}=C_{113}~,~~
    \mathcal{B}_{13}=-C_{112}~, \nonumber\\
    \mathcal{B}_{22}&= C_{213}~,~~
    \mathcal{B}_{23}=-C_{212}~,~~
    \mathcal{B}_{33}=C_{123}-C_{213}~,
\end{split}
\end{equation}
where we used that $C_{223}=-C_{113}$, $C_{312} = C_{213}-C_{123}$, $C_{313}=-C_{212}$ and $C_{323}=C_{112}$.
It is far more useful to decompose the rank-$2$ and rank-$3$ tensors by means of their irreducible representations of SO($3$). 
Following Ref.~\cite{Poisson:2009qj}, we introduce the radial unit vector $\Omega^i \equiv x^i/r$, with $r = \sqrt{\delta_{ij}x^i x^j}$ being the Euclidean radius representing the distance from the geodesic. The projector to the space orthogonal to $\Omega^i$ is given by $\gamma^{ij} = \delta^{ij} - \Omega^i \Omega^j$.
The electric quadrupole moment $\mathcal{E}_{ij}$ decomposes as follows
\begin{equation}\label{eq: dec}
    \mathcal{E}_{ij}=\mathcal{E}^{\text{q}} \left(\Omega_i \Omega_j -\frac{1}{2} \gamma_{ij} \right) + 2\mathcal{E}^{\text{q}}_{(i}\Omega^{}_{j)}+\frac{1}{2}\mathcal{E}^{\text{q}}_{\langle i j \rangle}~,
\end{equation}
where the scalar $\mathcal{E}^{\text{q}}$, the transverse vector $\mathcal{E}^{\text{q}}_{i}$ (\emph{i.e.} $\Omega^i  \mathcal{E}^{\text{q}}_{i} =0$) and the transverse STF tensor $\mathcal{E}^{\text{q}}_{\langle i j \rangle}$ are given by
\begin{align}\label{Edec}
    \nonumber
    \mathcal{E}^{\text{q}} &\equiv ~ \Omega^i \mathcal{E}_{ij} \Omega^j = -\gamma^{ij} \mathcal{E}_{ij}~,
    \\
    \mathcal{E}^{\text{q}}_{i} &\equiv \gamma^{\;j}_i \mathcal{E}_{jk} \Omega^k~,
    \\ \nonumber
    \mathcal{E}^{\text{q}}_{\langle i j \rangle} &\equiv 2\gamma^{\;k}_i\gamma^{\;l}_j \mathcal{E}_{kl}-\mathcal{E}_{kl}\gamma^{kl}\gamma_{ij}=2\gamma^{\;k}_i\gamma^{\;l}_j \mathcal{E}_{kl}+\mathcal{E}^q\gamma_{ij}~.
\end{align}
Similarly, for the magnetic quadrupole moment $\mathcal{B}_{ij}$, one has~\footnote{We used the decomposition of the rank-$3$ tidal tensor 
\begin{equation}
    C_{ijk} =\mathcal{B}^{\text{q}}_k \left( \Omega_i \Omega_j -\gamma_{ij} \right) - \mathcal{B}^{\text{q}}_j \left(\Omega_i\Omega_k  - \gamma_{ik}\right) +\frac{1}{2}\left(\mathcal{B}^{\text{q}}_{\langle ik\rangle}\Omega_j-\mathcal{B}^{\text{q}}_{\langle ij\rangle}\Omega_k\right)~,
\end{equation}
with the inverse relations given by
\begin{equation}
\mathcal{B}_i^q=C_{jki}\Omega^j\Omega^k~,~~~~\mathcal{B}^q_{\langle ij\rangle}=2\Omega^kC_{lk(i}\gamma^l_{~j)}~.
\end{equation}}
\begin{equation} \label{eq: dec2}
    \mathcal{B}_{ij} = \epsilon^{lk}_{~~(i} \left[\mathcal{B}^{\text{q}}_l \left( \Omega_{j)}\Omega_k - \gamma_{j)k} \right) + \frac{1}{4} \left(\mathcal{B}^{\text{q}}_{\langle j) l\rangle} \Omega_k - \mathcal{B}^{\text{q}}_{\langle j) k\rangle} \Omega_l \right) \right]~,
\end{equation}
with symmetrization with respect to the indices $(i,j)$ and STF with respect to the indices $\langle jl\rangle$ and $\langle jk\rangle$. The transverse vector $\mathcal{B}^{\text{q}}_i$ and the transverse STF tensor $\mathcal{B}^{\text{q}}_{\langle i j \rangle}$ are 
\begin{align}\label{Bdec}
\begin{split}
    \mathcal{B}^{\text{q}}_{i} &\equiv \epsilon_{ijk} \Omega^j\mathcal{B}^k_{~l}\Omega^l~,\\
    \mathcal{B}^{\text{q}}_{\langle i j \rangle} &\equiv 2\epsilon_{kl(i}\gamma^{m}_{~j)}\Omega^k\mathcal{B}^{l}_{~m}~.
\end{split}
\end{align}

\section{Hierarchical triple system}
\label{Sec:triplete}

In this section we apply the formalism introduced in Sec.~\ref{sec: tidal construction} to
an EMR binary system moving in the background of a Kerr black hole.
The EMR binary system consists of a Schwarzschild black hole of mass $M$ along with a test-particle of mass $m\ll M$.
We assume that the  black hole 
with mass $M_*$ moves slowly relatively to the EMR binary system $(M, m)$ and that one can describe the effect on the binary system to a good approximation by taking into account only the quadrupole tidal moments induced by $M_*$. This is valid provided
\begin{equation}
\label{smalltide}
    M^2 \ll \frac{\hat{r}^3}{M+M_*} \,,
\end{equation}
where $\hat{r}$ is the Boyer-Lindquist radius at which $M$ is located in the Kerr spacetime geometry induced by $M_*$~\cite{Poisson:2009qj}.
This arises from having two widely separated scales: one scale is the length scale of the Schwarzschild black hole $M$, the other is the curvature length scale $\mathcal{R}$ induced by the Kerr black hole $M_*$ at the location of $M$. We then require $M \ll \mathcal{R}$. This is called {\it small-tide approximation}~\cite{Poisson:2009qj} and it makes it possible to describe the motion of the binary system $(M, m)$ in the external Kerr geometry, ensuring that the tidal deformation is weak.
We can therefore describe the influence of $M_*$ on the binary system $(M, m)$ using, to a first approximation, the quadrupole tidal moments induced by the Kerr black hole itself. Since  $\mathcal{R}\sim \sqrt{\hat{r}^3/(M+M_*)}$ this, combined with the condition $M \ll \mathcal{R}$, gives the condition \eqref{smalltide}.
One natural way to achieve the condition \eqref{smalltide} is that $M$ is  much smaller than $M_*$, here called the \emph{hierarchical regime}
\begin{equation}
\label{hierar_regime}
M \ll M_* \,.
\end{equation}
This implies \eqref{smalltide} since $\hat{r} \gtrsim M_*$. In this case we have a hierarchical triple system of black holes $m \ll M \ll M_*$ (note that one could imagine both $M$ and $M_*$ being a supermassive black hole, but with a mass hierarchy). The hierarchical triple system is the case that we shall primarily consider in this paper, since the dynamics of the triple system in general will depend on the full expressions of the quadrupole tidal moments of the Kerr black hole $M_*$.

Another way to achieve the condition \eqref{smalltide} is the case where $M$ and $M_*$ are widely separated, here called the \emph{weak field regime}
\begin{equation}
\label{weakfield_regime}
M_* \ll \hat{r}~,
\end{equation}
assuming as well that $M \lesssim M_*$. This means one can consider two  black holes $M$ and $M_*$ of similar magnitude.
In this case the expression of the tidal moments induced by the Kerr black hole simplifies considerably~\cite{Yang:2017aht}
due to the fact that frame-dragging effects induced by the Kerr black hole can be neglected (see discussion around and below Eq.~\eqref{etageneral} for further detail).

It is also important to consider the time scales involved in our approximation. 
For simplicity, we consider the binary system having an orbit of $m$ around the Schwarzschild black hole of mass $M$ such that $r = \mathcal{O}(M)$. Then the time scale of the binary system is simply $\tau_{\rm binary} = \mathcal{O}(M)$.
Assuming $\hat{r} = \mathcal{O} (M_*)$
the time-scale associated with the orbit around the Kerr black hole of mass $M_*$ is $\tau_{\rm kerr} = \mathcal{O} ( M_*) $. Indeed, one can see explicitly from Eq.~\eqref{general_psi_dot} that we have $\dot{\Psi} = \mathcal{O}(1/M_*)$, which sets the rate of change of the angle $\Psi$. Thus, in the hierarchical regime \eqref{hierar_regime}, we have $\tau_{\rm kerr} \gg \tau_{\rm binary}$, which means that we can assume that the quadrupole moments and $\Psi$ do not vary with time.\\
Moreover, in the weak field regime \eqref{weakfield_regime}, the time scale for the orbit around the Kerr black hole is even larger $\tau_{\rm kerr} \gg M_*$ as the velocity will be non-relativistic. Thus, even if $M$ is of same order as $M_*$, we find that $\tau_{\rm kerr} \gg \tau_{\rm binary}$, and we can again neglect the time dependence of $\Psi$ and of the quadrupole moments.%
\footnote{A more general analysis can also take into account the regime $M \ll r \ll \mathcal{R}$ for which $\tau_{\rm binary} = \mathcal{O}(\sqrt{r^{3}/M})$.}
%

\subsection{Tidally deformed Schwarzschild spacetime}
We can describe the black hole with mass $M$ in the binary system using the tidally deformed Schwarzschild metric~\cite{Poisson:2009qj}.
Concretely, we add to the background metric $\bar{g}_{\mu\nu}$ a tidal perturbation $h_{\mu\nu}$
\begin{equation}\label{quadrupole_metric}
    ds^2 = \bar{g}_{\mu\nu}dx^{\mu}dx^{\nu} + h_{\mu\nu}dx^{\mu}dx^{\nu}~,
\end{equation}
where the tidal perturbation $h_{\mu\nu}$ is computed up to the first order in the small-tide approximation.
The background geometry (in spherical coordinates) is
\begin{align} 
    \bar{g}_{\mu\nu}dx^{\mu}dx^{\nu}&= - f dt^2 + \frac{dr^2}{f} + r^2 \Omega_{AB} d\theta^A d\theta^B~,
\end{align}
with $f=1-{2M}/{r}$ and $M$ being the black hole mass, $\theta^A=(\theta,\phi)$ and $\Omega_{AB}d\theta^A d\theta^B = d\theta^2 + \sin^2 \theta d\phi^2$ being the metric of the unit sphere. By only retaining the quadrupole order terms in the tidal deformation $h_{\mu\nu}$, one gets
\begin{equation}
\begin{split}
    &h_{\mu\nu}dx^{\mu}dx^{\nu}=
    \\
    &-  r^2\mathcal{E}^{\text{q}} \left( f dt+ dr \right)^2  - \frac{4}{3} r^3  \left( \mathcal{E}^{\text{q}}_A - \mathcal{B}^{\text{q}}_A\right) \left( f dt+dr \right) d\theta^A 
    \\
    &- \frac{1}{3} r^4 \left[ \left( 1 - \frac{2M^2}{r^2} \right) \mathcal{E}^{\text{q}}_{AB} -  \left( 1 - \frac{6M^2}{r^2} \right) \mathcal{B}^{\text{q}}_{AB} \right]  d\theta^A d\theta^B.
\end{split}
\end{equation}
The quadrupole moments are decomposed into the scalar $\mathcal{E}^{\text{q}}$, vector $\mathcal{E}^{\text{q}}_A$, $\mathcal{B}^{\text{q}}_A$ and tensor $\mathcal{E}^{\text{q}}_{AB}$, $\mathcal{B}^{\text{q}}_{AB}$ components, following the decomposition in Eqs.~\eqref{eq: dec}-\eqref{eq: dec2}, and are written in spherical coordinates.~\footnote{
For the sake of completeness, we write the change of coordinates from Cartesian to spherical coordinates:
\begin{equation*}
\label{eq: cartosph}
\begin{split}
    \mathcal{E}^{\text{q}}_i dx^i&=\frac{\partial x^i}{\partial x^A}\mathcal{E}^{\text{q}}_i dx^A=\mathcal{E}^{\text{q}}_\theta (rd\theta)+\mathcal{E}^{\text{q}}_\phi (rd\phi)~,
    \\
    \mathcal{E}^{\text{q}}_{\langle ij \rangle}dx^i dx^j&=\frac{\partial x^i}{\partial x^A}\frac{\partial x^j}{\partial x^B}\mathcal{E}^{\text{q}}_{\langle ij \rangle}dx^A dx^B=
    \\&=\mathcal{E}^{\text{q}}_{\theta\theta} (rd\theta)^2+2\mathcal{E}^{\text{q}}_{\theta\phi} r^2d\theta d\phi+\mathcal{E}^{\text{q}}_{\phi\phi} (rd\phi)^2~.
\end{split}
\end{equation*}
Similar considerations apply to the magnetic multipole moments $\mathcal{B}^{\text{q}}_{i}$ and $\mathcal{B}^{\text{q}}_{\langle ij\rangle }$.}
\begin{figure}[h!] 
    {\includegraphics[scale=0.095]{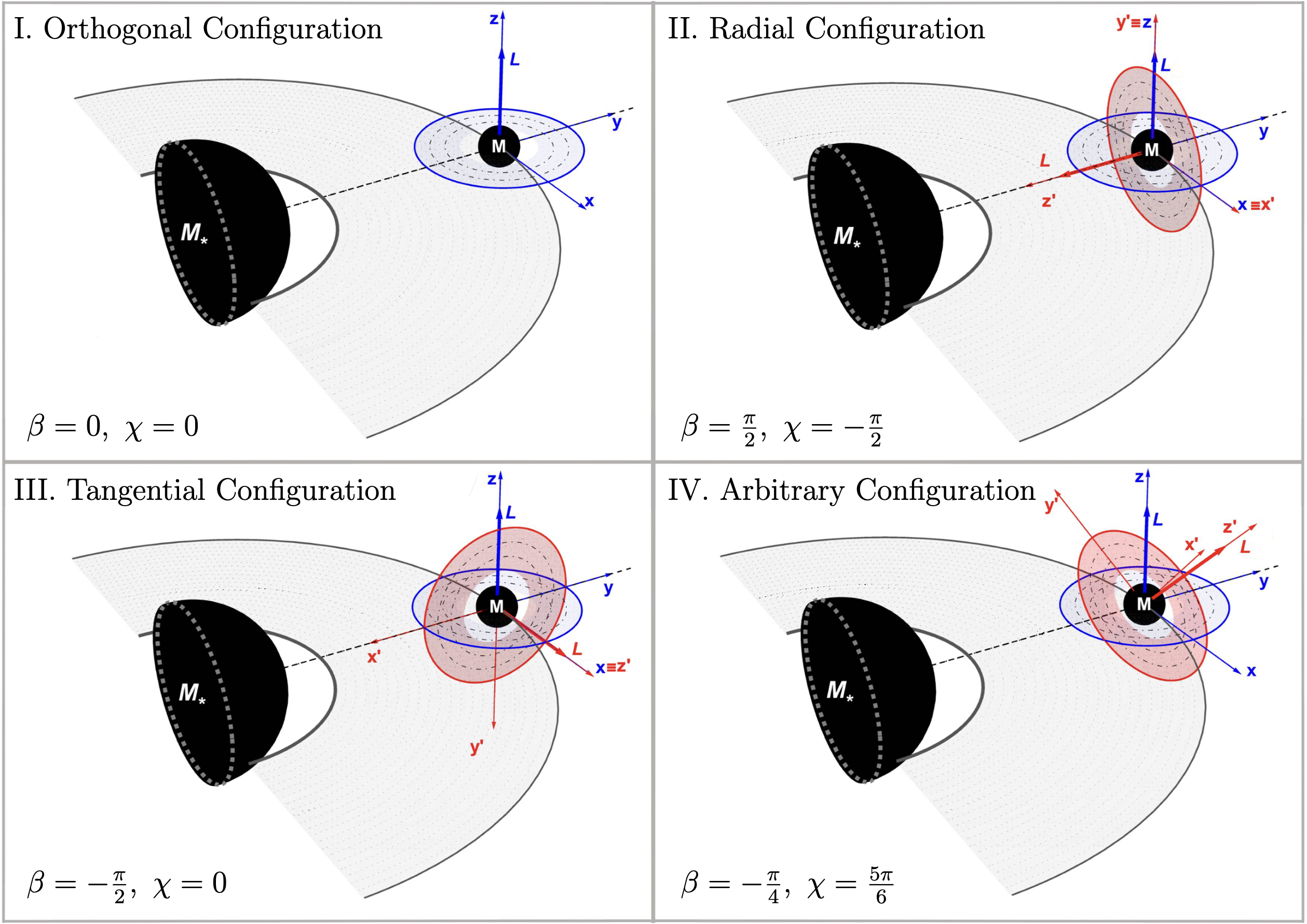}}
    \caption{
    For illustrative purposes, we show four possible configurations for a hierarchical three-body system $M_*\gg M\gg m$ in the special case for which the perturber $M_*$ is a Schwarzschild black hole and the EMR binary system $(M,m)$ is parallel-transported around a circular geodesic around $M_*$, whose orbital plane is depicted in gray and terminates at the ISCO. 
    These configurations are altered significantly in more general cases with a Kerr perturber or non-circular geodesics.
    The names of the configurations refer to the orientation of the orbital angular momentum $L$ of the binary system with respect to the gray orbital plane. The grey curve represents the orbit around $M_*$. The blue orbit marks a conventional ``initial" orthogonal configuration for the binary system reference frame, with the Cartesian axis oriented according to the parallel transported tetrad (panel I). The red orbits in panels II, III and IV are obtained by Euler rotations with angles written in the bottom-left of each panel.}
    \label{fig:4configurations}
\end{figure}
For an accurate description of our triple system, it is useful to identify the relative orientation between the orbital plane of the 
Kerr black hole -- responsible for the tidal deformation -- and the orbital plane where the dynamics of the EMR binary system $(M, m)$ takes place; see Fig.~\ref{fig:4configurations} illustrating four possible configurations in the special case when $M_*$ is a Schwarzschild black hole and the binary system is moving on a circular geodesic.
To describe an arbitrary configuration, one first introduces the unit directional vector
\begin{equation}
    \label{eq: Omega_0}
    \Omega^i=(\cos\phi \sin\theta, \sin\phi \sin\theta, \cos\theta)\;,
\end{equation}
centered in the Schwarzschild black hole of mass $M$, and attached to the reference frame of the EMR system $(M, m)$. 
One then sets, without loss of generality, the polar angle in the Schwarzschild reference system $\theta=\pi/2$: this is because the orbital motion takes place on an orbital plane and we set it to be the equatorial plane.  
Any arbitrary orientation is therefore given by performing a rotation on the unit vector in Eq.~\eqref{eq: Omega_0}, namely,
\begin{equation} 
    \label{Omegabeta}
    \vec{\Omega}'=R_\chi R_\beta R_\alpha \cdot \vec{\Omega}~, 
\end{equation}
with the Euler rotational matrices 
\begin{align}
\begin{split}
R_\alpha &=
    \begin{pmatrix}
    \cos\alpha &\sin\alpha&0 
    \\
    -\sin\alpha&\cos\alpha&0
    \\
    0&0&1
    \end{pmatrix}
~,~  
    R_\beta =
    \begin{pmatrix}
    1 &0&0 
    \\
    0&\cos\beta&\sin\beta
    \\
    0&-\sin\beta &\cos\beta
    \end{pmatrix}
~,\\ 
& \qquad\qquad\quad~ R_\chi =
    \begin{pmatrix}
    \cos\chi &\sin\chi&0 
    \\
    -\sin\chi&\cos\chi&0
    \\
    0&0&1
    \end{pmatrix}
~.
\end{split}
\end{align}
Note that Eq.~\eqref{Omegabeta} is only one among the 12 possible permutations of Euler matrices.
Furthermore, since we aim at describing a circular equatorial orbit in the binary system, it turns out that one of the Euler angle -- $\alpha$ in our convention -- can always be reabsorbed by a redefinition of the Schwarzschild azimuthal angle $\phi\to \phi+\alpha$. As a consequence, any orientation of a Schwarzschild orbit with respect to the Kerr perturber is specified only by the two angles $\beta$ and $\chi$.

\subsection{Tidal moments in spherical coordinates}
\label{sphericaltidal}

The tidal moments also depend on the relative configuration between the binary system $(M, m)$ and the Kerr perturber. Here, we compute the explicit expression of the tidal quadrupole moments associated to an arbitrary configuration. We recall that we set $\theta=\pi/2$ because we start with an equatorial orbit around the Schwarzschild black hole.
In Fig.~\ref{fig:4configurations} we have illustrated this and other configurations obtained by Euler rotations in the special case for which $M_*$ is a Schwarzschild black hole and the binary system moves on a circular geodesic.
In spherical coordinates, the decomposition of the electric quadrupole moment in its scalar, transverse vector and STF tensor components is given by Eq.~\eqref{Edec}, where the unit directional vector $\Omega^i$ is now replaced by the more general $\Omega'^i$ defined in Eq.~\eqref{Omegabeta}. 
The electric quadrupole moments read as
\begin{align}
\label{general E}
    \begin{split}
        \mathcal{E}^{\text{q}} &= -\frac{1}{8}\left(C_{33}+\mathcal{T}^+_2+\mathcal{T}^+_4-4 \mathcal{T}^+_3 \sin 2\phi  \right) 
        \\
        &\quad-\frac{1}{8} \left[3(C_{33}+\mathcal{T}^+_2)-\mathcal{T}^+_4\right]\cos 2\phi ~,\\
        \mathcal{E}^{\text{q}}_{\theta} &= \frac{1}{4}\left[2 \mathcal{T}^-_3 \cos \phi - \mathcal{T}^-_4 \sin \phi \right]~,\\
        \mathcal{E}^{\text{q}}_{\phi} &= \frac{1}{8}\left[4 \mathcal{T}^+_3 \cos 2\phi + \left(3(C_{33}+\mathcal{T}^+_2)-\mathcal{T}^+_4\right)\sin 2\phi \right]~,\\
        \mathcal{E}^{\text{q}}_{\theta\theta} &= -\mathcal{E}^{\text{q}}_{\phi\phi} = \mathcal{E}^{\text{q}}+\frac{1}{2}\left(C_{33}+\mathcal{T}^+_2+\mathcal{T}^+_4\right)~,\\
        \mathcal{E}^{\text{q}}_{\theta\phi} &= -\frac{1}{2} \left(2 \mathcal{T}^-_3 \sin \phi + \mathcal{T}^-_4 \cos \phi \right)~,
    \end{split}
\end{align}
where we defined the following rotations around $\chi$ of the components of $C_{ij}$ 
    \begin{align}
    \begin{split}
        \mathcal{T}^+_1 &= C_{23}\cos \chi + C_{13}\sin \chi~,
        \\ 
        \mathcal{T}^-_1 &= C_{23}\sin \chi - C_{13}\cos \chi~, \\
        \mathcal{T}^+_2 &= 2C_{12}\sin 2\chi + (2C_{22}+C_{33})\cos 2\chi ~, \\ 
        \mathcal{T}^-_2 &=2C_{12}\cos 2\chi - (2C_{22}+C_{33})\sin 2\chi 
        \end{split}
    \end{align}
and the rotations around $\beta$ of $\mathcal{T}^{\pm}_{1,2}$
    \begin{align}
    \begin{split}
        \mathcal{T}^+_3 &= 2\mathcal{T}^-_1 \sin \beta + \mathcal{T}^-_2 \cos \beta~,\\ 
        \mathcal{T}^-_3 &= 2\mathcal{T}^-_1 \cos \beta - \mathcal{T}^-_2 \sin \beta~,\\
        \mathcal{T}^+_4 &= 4 \mathcal{T}^+_1 \sin 2\beta + (3C_{33}-\mathcal{T}^+_2)\cos 2\beta~,\\ 
        \mathcal{T}^-_4 &= 4 \mathcal{T}^+_1 \cos 2\beta - (3C_{33}-\mathcal{T}^+_2)\sin 2\beta \,.
    \end{split}
    \end{align}
Similarly for the magnetic quadrupole moments, whose decomposition is given in Eq.~\eqref{Bdec}, we find that 
    \begin{align}\label{generalB}
    \begin{split}
        \mathcal{B}^{\text{q}}_{\theta}&= \frac{1}{8}\left[4 \mathcal{S}^+_3 \cos 2\phi + \left(3(C_{312}-\mathcal{S}^+_2) -\mathcal{S}^+_4\right)\sin 2\phi\right]~,\\
        \mathcal{B}^{\text{q}}_{\phi}&= -\frac{1}{4}\left(2\mathcal{S}^-_3 \cos \phi -\mathcal{S}^-_4 \sin \phi\right)~,\\
        \mathcal{B}^{\text{q}}_{\theta\theta}&= -\mathcal{B}^{\text{q}}_{\phi\phi}=-\frac{1}{2}\left(2\mathcal{S}^-_3 \sin \phi +\mathcal{S}^-_4 \cos \phi\right)~,\\
        \mathcal{B}^{\text{q}}_{\theta\phi}&= -\frac{3}{8} \left(C_{312} -\mathcal{S}^+_2 +\mathcal{S}^+_4+\frac{4}{3} \mathcal{S}^+_3 \sin 2\phi\right)
        \\
        &\quad+\frac{1}{8}\left[ 3(C_{312}-\mathcal{S}^+_2) -\mathcal{S}^+_4\right]\cos 2\phi~,
    \end{split}
    \end{align}
where we defined the rotations around $\chi$ of the components of $C_{ijk}$
    \begin{align}
    \begin{split}
        \mathcal{S}^+_1 &= C_{212}\cos \chi + C_{112}\sin \chi~,\\ 
        \mathcal{S}^-_1 &= C_{212}\sin \chi - C_{112}\cos \chi~,\\
        \mathcal{S}^+_2 &= 2C_{113}\sin 2\chi + (C_{123}+C_{213})\cos 2\chi ~, \\ 
        \mathcal{S}^-_2 &=2C_{113}\cos 2\chi - (C_{123}+C_{213})\sin 2\chi
    \end{split}
    \end{align}
and the rotations around $\beta$ of $\mathcal{S}^{\pm}_{1,2}$
    \begin{align}
    \begin{split}
        \mathcal{S}^+_3 &= 2\mathcal{S}^-_1 \sin \beta - \mathcal{S}^-_2 \cos \beta~,\\ 
        \mathcal{S}^-_3 &= 2\mathcal{S}^-_1 \cos \beta + \mathcal{S}^-_2 \sin \beta~,\\
        \mathcal{S}^+_4 &= 4 \mathcal{S}^+_1 \sin 2\beta + (3C_{312}+\mathcal{S}^+_2)\cos 2\beta~,\\ 
        \mathcal{S}^-_4 &= 4 \mathcal{S}^+_1 \cos 2\beta - (3C_{312}+\mathcal{S}^+_2)\sin 2\beta~.
    \end{split}
    \end{align}
The structure of the tidal quadrupole moments \eqref{general E} and \eqref{generalB} is the following: the tidal deformations sourced by a generic third body over the EMR binary system $(M, m)$ are fully encoded in the tidal tensors $C_{ij}$ and $C_{ijk}$, while the angles $\beta$ and $\chi$, parametrizing the relative orientation between the third body and the binary system, affect the tidal effects over the binary system. 
We remark that the above expressions of the tidal quadrupole moments are general, and can also be employed to model environmental effects in numerical works.
In the specific case of a Kerr black hole as a third body responsible for the tidal deformations, the explicit expressions of the tidal tensors $C_{ij}$ and $C_{ijk}$ are given, respectively, in Eqs.~\eqref{Cij_Kerr} and \eqref{Cijk_Kerr}.
We anticipate here another property of the tidal quadrupole moments. As we shall see in the next section, it is often useful to define the secular average over the azimuthal angle $\phi$. The explicit dependence of the tidal quadrupole moments \eqref{general E} and \eqref{generalB} implies that only $\mathcal{E}^{\text{q}}$ (and $\mathcal{E}^{\text{q}}_{\theta\theta} = -\mathcal{E}^{\text{q}}_{\phi\phi}$) as well as $\mathcal{B}^{\text{q}}_{\theta\phi}$ are relevant for physical observables upon secular averaging.

\section{Secular dynamics of binary system}
\label{sec: secular H}

In this section we focus on the secular dynamics of the binary system $(M,m)$, \emph{i.e.} the dynamics of the binary system  after a large number of orbits of the test particle of mass $m$, and analyze how it is affected by the tidal fields induced by the Kerr perturber of mass $M_*$, in the hierarchical regime $m \ll M \ll M_*$.
More specifically our goal is to understand how the orbital parameters of the test particle around the Schwarzschild black hole, such as the energy or the angular momentum, are shifted by the presence of an external tidal field.
%

\subsection{Secular Hamiltonian of test particle in binary system}

Following the setup of the previous section, we focus on the orbital motion of the object of mass $m$, approximated as a test particle, taking place on the equatorial plane of the Schwarzschild black hole. This amounts to set $\theta = \pi/2$.
We approximate the four-velocity as
\begin{equation}
\label{eq: p-epsilon}
    u^\mu \simeq \bar{u}^\mu + \, u_{(1)}^\mu \,,
\end{equation}
where $\bar{u}^\mu$ is the 4-velocity of the unperturbed bound orbit, which can be taken as circular or elliptic, and $u_{(1)}^\mu$ is the leading correction due to the tidal perturbation $h_{\mu\nu}$. 
In this work, we focus on perturbations of circular orbits $\bar{u}^\mu = (\bar{E}/f,0,0,\bar{L}/r^2)$ on the Schwarzschild background metric $\bar{g}_{\mu\nu}$. Here $\bar E = -\bar{u}^{\mu}\bar{g}_{\mu\nu}(\partial_t)^{\nu}$ and $\bar L=\bar{u}^{\mu}\bar{g}_{\mu\nu}(\partial_{\phi})^{\nu}$ are the conserved energy and angular momentum of the test particle in the Schwarzschild geometry.
Tidal deformations to the four-velocity affect the gauge-independent photon red-shift measurements \cite{Detweiler:2008ft} ($\sim u^t_{(1)}$), are responsible for radial deviations ($\sim u^r_{(1)}$), tilt the orbital plane ($\sim u^{\theta}_{(1)}$), and shift the orbital frequency ($\sim u^\phi_{(1)}$).
The Hamiltonian of a test particle moving around a tidally deformed Schwarzschild black hole (see Eq.~\eqref{quadrupole_metric}) is given by
\begin{equation}
    \label{eq: H0}
    H=\frac{1}{2}u^\mu u^\nu g_{\mu\nu} \simeq \frac{1}{2} \bar{u}^{\mu}\left(\bar{u}^{\nu}+2 u_{(1)}^{\mu}\right)\bar{g}_{\mu\nu} + \frac{1}{2} \bar{u}^{\mu}\bar{u}^{\nu}h_{\mu\nu} \,  .
\end{equation}
In the specific case of a circular orbit $\bar{u}^{\mu}$ in the Schwarzschild background metric $\bar{g}_{\mu\nu}$, radial and polar deviations affects the dynamics only at higher order \cite{Yang:2017aht,Isoyama:2014mja}. Moreover, from Eq.~\eqref{eq: H0}, the tidal perturbations that enter the Hamiltonian are $h_{tt} \propto \mathcal{E}^{\text{q}}$, $h_{t\phi} \propto \mathcal{E}^{\text{q}}_{\phi},\mathcal{B}^{\text{q}}_{\phi}$, and $h_{\phi\phi} \propto \mathcal{E}^{\text{q}}_{\phi\phi}, \mathcal{B}^{\text{q}}_{\phi\phi}$.
A further simplification, that is very common in celestial mechanics, is the secular averaging over a timescale of order of the orbital timescale. In general, secular averaging consists in integrating out the short-term oscillations (due to tidal effects) from the dynamics. Therefore, in our case, we integrate out characteristic timescales up to the orbital period, $\sqrt{r^3/M}$, of the test particle by averaging over its orbit.
From a geometrical perspective \cite{Yang:2017aht} the secular average can be understood by considering that
the effective dynamics of a test particle, which follows a tidally-deformed geodesic $\gamma'$ at the first order in $h_{\mu\nu}$, can be well captured by replacing the physical trajectory $\gamma'$ with an averaged circular trajectory $\gamma$ in the perturbed spacetime. 
The averaged geodesic $\gamma$ can be interpreted as a \emph{secular orbit} in the tidally perturbed background. 
We introduce the secular average of a quantity $\mathcal{A}$ as 
\begin{equation}
\label{eq:averagephi}
    \langle \mathcal{A} \rangle = \frac{1}{2 \pi} \int_{0}^{2\pi} \mathcal{A}\big|_{\gamma}~d\phi~,
\end{equation}
where $\phi$ is the azimuthal angle of the orbit and $\gamma$ is the averaged circular orbit on $g_{\mu \nu}$. In particular, if $\gamma'$ is quasi-circular, the averaged secular geodesic $\gamma$ deviates from the physical orbit $\gamma'$ only starting from second order in $h_{\mu\nu}$ in the Hamiltonian \eqref{eq: H0}.

After averaging, from Eqs.~\eqref{general E} and \eqref{generalB}, we get~\footnote{Our result differs from the one in Ref.~\cite{Yang:2017aht} where 
 $\langle h_{t\phi} \rangle \neq 0$.}
\begin{equation}
\begin{split}
    \langle h_{tt} \rangle &= -r^2 f^2 \langle \mathcal{E}^{\text{q}} \rangle~,\\
    \langle h_{t\phi} \rangle &= 0~,\\
    \langle h_{\phi\phi} \rangle &= -r^4\left(1-2\frac{M^2}{r^2}\right)\langle \mathcal{E}^{\text{q}} \rangle ~,
\end{split}
\end{equation}
and therefore the secular average of the Hamiltonian \eqref{eq: H0} up to quadrupole order can be recast as~\footnote{Notice that we used that $\langle u^{\mu}u^{\nu} g_{\mu\nu}\rangle\simeq\langle u^{\mu}\rangle \langle u^{\nu}\rangle \langle g_{\mu\nu}\rangle$ including corrections of order $h_{\mu\nu}$.}
\begin{equation}   
    \label{eq:averageH}
    \begin{split}
        \langle H \rangle  \simeq &-\frac{1}{2}\left(\frac{\langle E \rangle^2}{f} - \frac{\langle L \rangle^2}{r^2}\right) 
        \\
        &- \eta \left[\langle E \rangle^2 +\left(1-2\frac{M^2}{r^2}\right)\frac{\langle L \rangle^2}{r^2}\right]\frac{r^2}{M^2}~,
        \\
        \vspace{15mm}
    \end{split}
\end{equation}
where $\eta$ is a parameter that encodes all the effects of the tidal deformations at the quadrupole order. $E=-u^{\mu} g_{\mu\nu}(\partial_t)^{\nu}$ and $L=-u^{\mu} g_{\mu\nu}(\partial_{\phi})^{\nu}$ are, respectively, the energy and angular momentum with respect to the perturbed spacetime and the symbol $\langle \cdot \rangle$ stands for  secular average. 
We stress that $\langle E \rangle$ and $\langle L \rangle$ encode the kinematics (including the secular effects on the orbits), while the parameter $\eta$ effectively depends on the secular tidal deformations ($\propto C_{ij}$) and on the orientation $(\beta,\chi)$ of the binary system. More explicitly, we find that the tidal parameter $\eta$ is proportional to the secular average of the electric scalar tidal field
\\
\vspace{3mm}
\begin{widetext}
    \begin{equation} \label{etageneral}
\begin{split}
    \eta &= -\frac{M^2}{2} \langle \mathcal{E}^{\text{q}} \rangle \\
    &=\frac{M^2}{16} \bigg\{C_{33}\left(1+3\cos 2\beta\right) + 4\left(C_{13}\sin \chi + C_{23}\cos\chi\right)\sin 2\beta+ \left[2C_{12}\sin 2\chi + \left(2C_{22}+C_{33}\right)\cos 2\chi \right] \left(1-\cos 2\beta\right)  \bigg\}.
\end{split}
\end{equation}
Notice that this expression for $\eta$ can also be used for other tidal tensors $C_{ij}$ than the one induced by the Kerr black hole in this paper.
In fact, it is a general result for any EMR binary system consisting of a Schwarzschild black hole of mass $M$ and a test particle of mass $m$, under the assumptions that: 1) it is immersed in a tidal environment, 2) only the quadrupole order is retained and 3) the secular approximation is valid. 

If we specialize Eq.~\eqref{etageneral}  to the tidal tensors of a Kerr perturber that we presented in Sec.~\ref{sec: tidal construction} in Eq.~\eqref{Cij_Kerr}, it can be shown that the Marck's angle $\Psi$ appearing in the $C_{ij}$'s, which is a constant in this approximation, can be reabsorbed by a simple shift of the angle $\chi$ , $\chi\to\chi-\Psi$ so that $\eta$ is explicitly given by
\begin{align} \label{etaKerr}
\eta=\frac{I_1 M^2 }{16 K \Sigma ^2}&\left[3 ST(\hat{r}^2
-a^2  \cos^2 \hat{\theta})(1-4\sin^2\beta\sin^2\chi)+6 \cos 2 \beta  \left(\hat{r}^2 T^2-
   a^2 S^2 \cos ^2\hat{\theta}\right)
   \nonumber\right.
   \\
   &\left.~~-3 a \cos
   \hat{\theta} \left(a S^2 \cos
   \hat{\theta}+4 \hat{r} \sin 2 \beta 
   \sqrt{S T} (S+T) \sin
   \chi \right)+K\Sigma ^2+3 \hat{r}^2
   T^2\right]
   \\ 
   &\hspace{-1.4 cm}+\frac{3 I_2 M^2\sqrt{S T}}{4 K
   \Sigma ^2} \left[
 \left(a^2
   S \cos ^2\hat{\theta}-\hat{r}^2
   T\right)\sin 2 \beta \sin \chi  -2 a\hat{r}\sqrt{S T}\cos
   \hat{\theta}\left(\cos ^2\beta  -\sin^2\beta \sin^2\chi\right)\right],\nonumber
\end{align}
\end{widetext}
where $K$ is the Carter constant, and $I_1$, $I_2$, $S$ and $T$ are defined in Eqs.~\eqref{I1I2} and \eqref{ST}. 
In the weak field regime, where $ M_{\star} \ll \hat{r} $, the leading order part of $\eta$ is given by
\begin{equation}
\label{etaweakfieldgeneral}
\begin{split}
    \eta =& \frac{M^2}{4 K}\frac{M_{\star}}{\hat{r}^3} \bigg[ 3T(\cos^2\beta-\sin^2\beta\sin^2\chi)
    \\
    &-K\left(2-3\sin^ 2\beta\right) -3 a \sqrt{T}\cos \hat{\theta}  \sin \chi\sin 2\beta \bigg] ~.
\end{split}
\end{equation}
In  the equatorial plane of the Kerr perturber $\hat{\theta}=\pi/2$, the parameter $\eta$ takes the simpler form
\begin{equation} \label{etaweak}
    \eta = \frac{M^2}{4}\frac{M_{\star}}{\hat{r}^3} \left(1-3\sin^2\beta\sin^2\chi\right)~,
\end{equation}
that depends only on the two Euler angles $\chi$ and $\beta$ and not on the spin parameter $a$, so one cannot distinguish the effect of the tidal forces from the case of a Schwarzschild perturber ($a=0$). This is reasonable in the sense that if one goes at large distances on the equatorial plane, one cannot feel the effect of the spin of the Kerr black hole. 
For $\chi=\pi/2$, in particular, Eq.~\eqref{etaweak} coincides with the result of Ref.~\cite{Yang:2017aht}, provided one identifies $\beta$ as the angle between the tidal symmetry axis, parallel to $z$, and the orbital plane: $\eta=\frac{M^2 M_{\star}}{4\hat{r}^3}\left(1-3\sin^2\beta\right)$.

\subsection{Special case of circular equatorial geodesic in Kerr background}
\label{Sec:equatorial}
We emphasize that neither the construction of the tidal quadrupole moments in Sec.~\ref{sec: tidal construction}, nor the discussion about the secular dynamics of the Schwarzschild binary system in the current section rely on any assumption concerning the geodesic motion followed by the Schwarzschild black hole of mass $M$ around the Kerr black hole of mass $M_*\gg M$.
However, in order to simplify the discussion, we now focus on solutions of the geodesic equations \eqref{eq: Kerr_geod} describing circular ($\dot{\hat r}=0$) and equatorial geodesics ($\hat \theta=\pi/2$ and $\dot{\hat \theta}=0$) in the Kerr spacetime. Under these assumptions, the parameters that characterise the geodesic -- namely the energy, the angular momentum and the Carter's constant -- are written explicitly in Eq.~\eqref{eq: EL_Kerr}. In this case the effective parameter $\eta$ given in Eq.~\eqref{etaKerr} reduces to the simple expression 
\begin{equation}
\label{etaequatorial}
\begin{split}
    \eta = &\frac{M_*M^2}{16\hat{r}^3} \times\\
    &\times\left\{1+3\frac{K}{\hat{r}^2} -3 \left[\frac{K}{\hat{r}^2}+\left(1+\frac{K}{\hat{r}^2}\right)\sin^2\chi\right]\sin^2\beta\right\}~.
\end{split}
\end{equation}
Note that this is a general result, valid beyond the weak-field regime ($M_{\star} \ll \hat{r}$).

For a circular equatorial geodesic it is moreover easy to express the Carter constant $K$ in terms of the Kerr parameters $(a,M_*)$ and the orbital radius $\hat r$, by means of the following relation
\begin{equation}
    \frac{K}{\hat r^2}=-\frac{1}{2}\left(1-\frac{\hat r ^2-\hat r  M_*-2\sigma a \sqrt{\hat r  M_*} +2 a ^2}{\hat r ^2-3 \hat r  M_*+2\sigma a\sqrt{\hat r M_*}}\right)~.
\end{equation}
We recall that $\sigma=\pm1$ distinguishes whether a circular orbit is co-rotating or counter-rotating with respect to the Kerr black hole angular momentum.

An intriguing observation is that, from the expression \eqref{etaequatorial}, one can see that there exist certain configurations for the EMR binary system $(M,m)$ on the Kerr equatorial plane, such that $\eta=0$, namely such that the dynamical contribution of the tidal effects vanishes in the secular approximation.
For a given angle $\chi$, this holds when the angle $\beta=\beta^*(\chi)$ with
\begin{equation}
    \label{eq: beta_star}
    \sin^2\beta^*(\chi)=\frac{1+3 K/\hat r^2}{3\left[K/\hat r^2+ \left(1+ K/\hat r^2\right)\sin^2\chi\right]}~.
\end{equation}
In the weak-field limit this relation reduces to $\sin^2\beta^*(\chi)=(3\sin^2\chi)^{-1}$, thus generalising the result obtained in Ref.~\cite{Yang:2017aht}, which is valid only for $\chi=\pi/2$.
Instead, the above result goes beyond the weak-field regime, and can be used also for circular geodesics close to the event horizon of Kerr.

\begin{figure}
    \centering
    \includegraphics[scale=0.25]{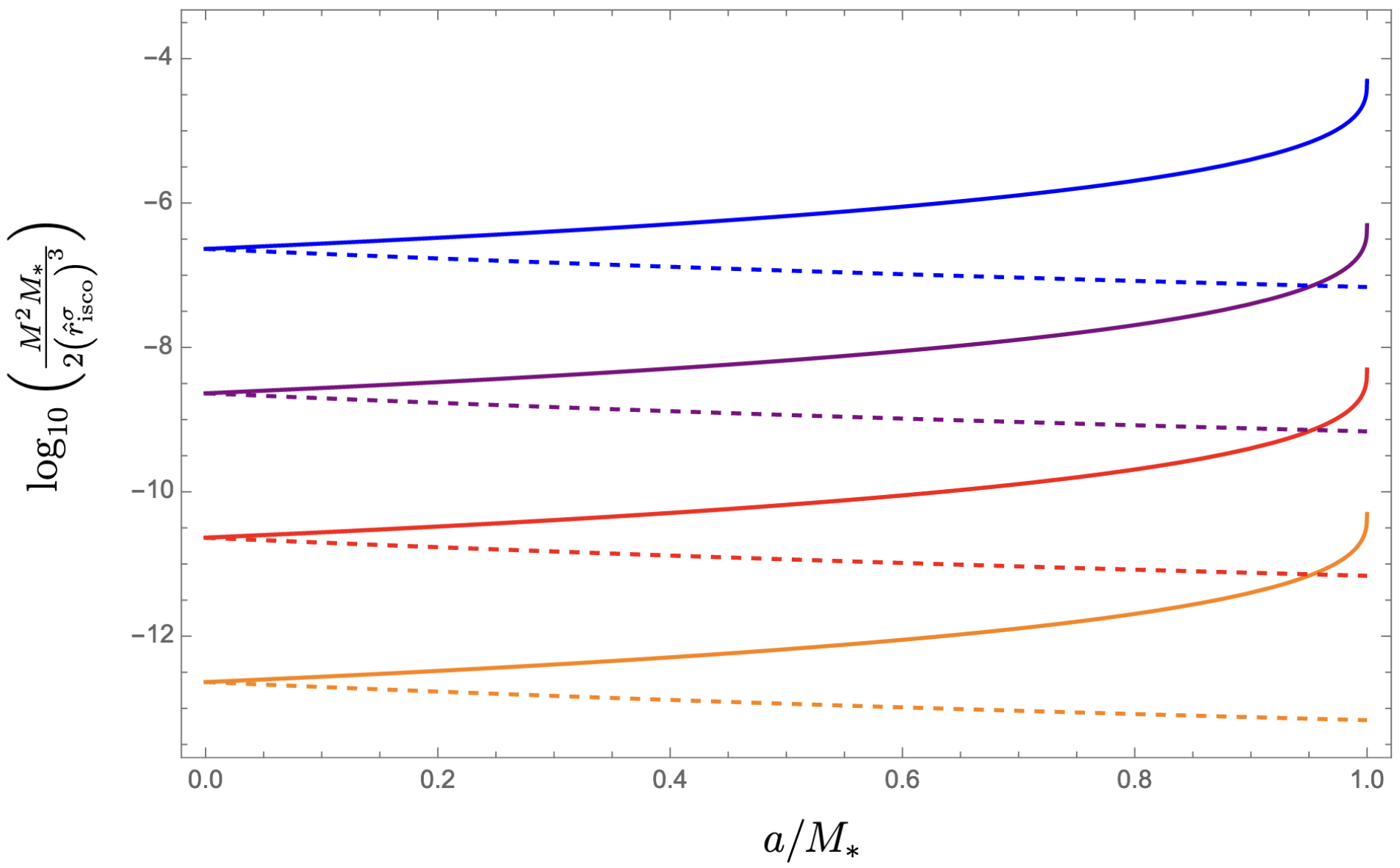}
    \caption{The picture represents how $\eta$, when evaluated at the ISCO $\hat r\equiv \hat r^\sigma_{\rm isco}$, depends on the black hole spin $a$.
    The logarithm of the prefactor in Eq.~\eqref{eq: eta_ISCO} is considered in order to have a clear distinction for the curves. Colors are used to represent different magnitudes for the ratio $\mu=M/M_*$. In particular $\mu= 10^{-2}$ in blue, $\mu= 10^{-3}$ in purple, $\mu= 10^{-4}$ in red and $\mu= 10^{-5}$ in orange. Solid lines are representative for the co-rotating ISCO $\sigma=1$, whereas dashed lines for counter-rotating ISCO $\sigma=-1$.}
    \label{fig:etaISCOspin}
\end{figure}

Among all the time-like equatorial circular orbits, the ISCO stands out for its relevance in black hole astrophysics. We recall that two ISCOs exist in the equatorial plane of a Kerr black hole, one which is co-rotating ($\sigma=+1$) and the other counter-rotating ($\sigma=-1$). 
In light of migration trap mechanisms that could lead to the formation of black hole binary systems \cite{Bellovary:2015ifg,Secunda:2020cdw,Peng:2021vzr}, it is interesting to analyse the case where the circular equatorial orbit, in which the binary system is located, is given by the Kerr ISCOs. 
More specifically, in the following we set $r \equiv \hat{r}^\sigma_{\rm ISCO}$, with
\begin{equation}
\label{eq: isco_Kerr}
    \hat{r}^\sigma_{\rm ISCO}= M_*\left[3+Z_2-\sigma \sqrt{(3-Z_1)(3+Z_1+2Z_2)}\right]~,
\end{equation}
where
\begin{equation}
\begin{split}
    Z_1 &=1+\left(1-\frac{a^2}{M_*^2}\right)^{\frac{1}{3}}\left[\left(1+\frac{a}{M_*}\right)^{\frac{1}{3}}+\left(1-\frac{a}{M_*}\right)^{\frac{1}{3}}\right]~,
    \\
    Z_2&=\sqrt{Z_1^2+3\frac{a^2}{M_*^2}}~.
\end{split}
\end{equation}
It is possible to show that the following relation implicitly defines the ISCOs in terms of the conserved Killing energy~\cite{Bardeen:1972fi}
\begin{equation}
\label{eq: implicitISCO_E}
    \hat{E}_{\rm ISCO}^2=1-\frac{2}{3}\frac{M_*}{\hat r^\sigma_{\rm ISCO} }~,
\end{equation}
so that, by combining the expression above with $K=(a \hat{E}-\hat{L})^2$ as in Eq.~\eqref{eq: EL_Kerr}, one obtains that the Carter constant at the ISCOs takes the value $K=1/3\left( \hat{r}^\sigma_{\rm ISCO}\right)^2$. The expression for $\eta$ in this limit considerably simplifies and it is given by
\begin{equation}
\label{eq: eta_ISCO}
    \eta=\frac{M^2M_*}{2\left( \hat{r}^\sigma_{\rm ISCO}\right)^3}\left[1-\frac{1}{2}(1+4\sin^2\chi)\sin^2\beta\right]~.
\end{equation}
Notice that, even if $\hat{r}^\sigma_{\rm ISCO}\sim\mathcal{O}(M_*)$, the small tide approximation Eq.~\eqref{smalltide} is still valid since $M\ll M_*$. 
This means that one can still legitimately consider the quadrupole approximation for a hierarchical three-body system in which the binary system ($M,m$) is orbiting on the ISCO of the Kerr black hole of mass $M_{\star}$.
It is interesting to notice that in the expression \eqref{eq: eta_ISCO} the dependence on the spin parameter of the Kerr perturber is only contained in the prefactor, whereas the part inside square brackets specifies the configuration of the binary system. A plot of the prefactor showing the dependence on the spin of the Kerr black hole is shown in Fig.~\ref{fig:etaISCOspin} for different values of the ratio $M/M_*$.

It is also interesting to observe that the expression for $\eta$ at the ISCO remains well-defined even when the Kerr black holes is rotating close to extremality, namely for $a\to M_*$. In this case one has $\hat{r}^+_{\rm ISCO}\to M_* $, so that the prefactor only depends on the ratio $M^2/M_*^2$. It is also evident by means of the plot in Fig.~\ref{fig:etaISCOspin} that the extreme case represents the maximum value of $\eta$ at the ISCO for a given configuration of the binary system.  

\begin{figure}
    \centering
    \includegraphics[scale=0.25]{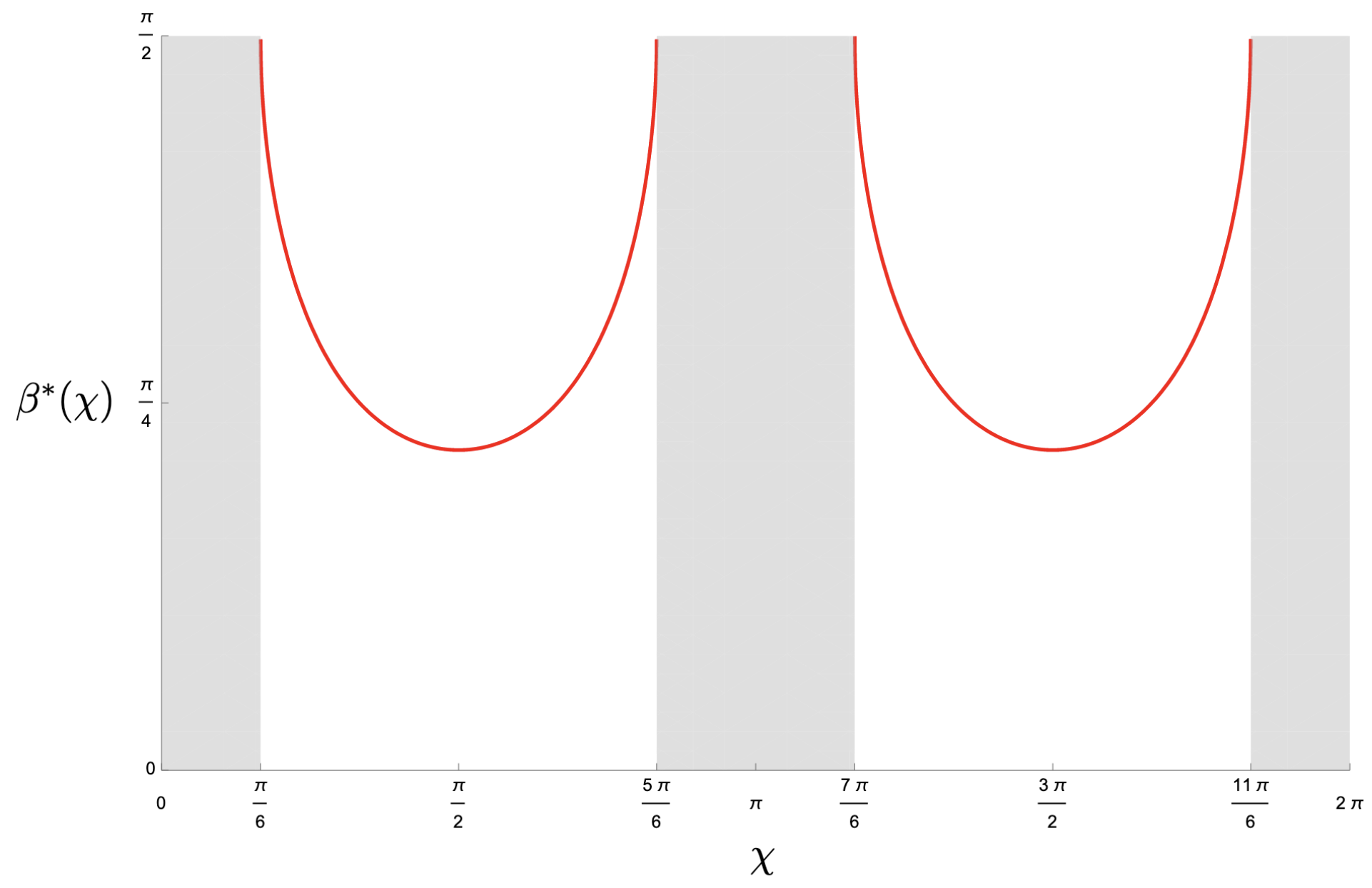}
    \vspace{3mm}
    \caption{The red line identifies the configurations $\beta^*(\chi)$ for which the secular effect of tidal deformations vanishes under the assumption $\hat r\equiv \hat  r^\sigma_{\rm ISCO}$. The gray areas represent exclusion zones, namely values of the angle $\chi$ in which the relation \eqref{eq: beta_star_ISCO} cannot be satisfied. More specifically, these corresponds to values of $\chi$ that would lead $|\sin^2\beta^*|>1$. }
    \label{fig: iscobeta*1}
\end{figure}

For the EMR binary system moving on the ISCO in the Kerr black hole spacetime, we can get the angle $\beta=\beta^*(\chi)$, as a function of the angle $\chi$, for which $\eta=0$, at which the tidal effects vanish from the secular dynamics of the binary system. Using that $K/(\hat r^{\sigma}_{\rm ISCO})^2=1/3$, one gets
\begin{equation}
    \label{eq: beta_star_ISCO}
     \sin^2 \beta^*(\chi)=\frac{2}{1+4\sin^2\chi}~.
\end{equation}
In Fig.~\ref{fig: iscobeta*1} we show the admissible values of $\beta^*(\chi)$ when the binary system is at the ISCO.
%

\section{Secular shifts for ISCO and photon sphere}
\label{sec: secular shift}

In this section we investigate how the tidal deformations affect the secular motion of the characteristic orbits of a test-particle around a Schwarzschild black hole using the Hamiltonian given in Eq.~\eqref{eq:averageH}. In particular, we consider two specific orbits in the case of general configurations of the three-body system, namely the ISCO and the photon sphere in the perturbed Schwarzschild spacetime. Before computing tidal effects on the orbital motion, we address the issue of gauge invariance of such effects.

\subsection{Gauge invariance of secular observables}

We start by recalling that the energy $E$ can be expressed in terms of the Killing vector $\partial_t$, namely 
\begin{equation}
E = -  u^\mu g_{\mu\nu}T^\nu \,,
\end{equation}
where in our coordinates $T = \partial_t$ and 
$g_{\mu\nu}$ and $u^\nu$ are the metric and four-velocity including tidal perturbations.
Given that $T$ is a Killing vector field, $dE/d\tau=0$ in any coordinate system when evaluated on a geodesic.
Therefore, $E$ is conserved and gauge invariant.

The angular momentum can be covariantly written as 
\begin{equation}
L = u^\mu g_{\mu\nu} J^\nu \,,
\end{equation}
where in our coordinates $J=\partial_{\phi}$.
However, as $J$ is not a Killing vector field for the full metric $g_{\mu \nu}$, $L$ is not conserved along geodesics. The strategy here is to get a conserved quantity and show that it is also gauge invariant.
We assume that the angular momentum $L$ can be expanded as 
\begin{equation}
L \simeq \bar{L} + \eta L_1 \,,
\end{equation}
where $\bar{L}$ is the conserved angular momentum in the Schwarzschild background, while $L_1$ is the correction induced by the tidal fields at the quadrupole order, which in general it is not conserved. 
The key observation is that the averaged metric field $\langle g_{\mu \nu}\rangle$ does not depend on $\phi = \phi(\tau)$, implying that $\langle L \rangle$ is now a conserved quantity along the secular geodesic.
Therefore, for a quasi-circular orbit we can write 
\begin{equation}
\label{eq:L}
\langle L \rangle \simeq\int_0^{2\pi} \left(\bar{L}+\eta L_1\right)|_{\gamma} d\phi  = 2\pi \bar{L} + \eta \int_0^{2\pi}L_1|_{\gamma} d\phi ~.
\end{equation}

We now consider a class of coordinate transformations which, up to the quadrupole order, acts on the rotation angle according to
\begin{equation}
\label{eq:gauge}
    \phi \to \tilde{\phi}\simeq \phi+\eta~\chi(r,\theta,\phi) ~,
\end{equation}
such that $\chi$ is a periodic function of $\phi$ with a period of $2 \pi$, namely $\chi (\phi)=\chi (\phi +2\pi)$, and the gauge variations of the coordinates $r$ and $\theta$ are of $\mathcal{O}(\eta)$. These assumptions guarantee that quasi-circular orbits are always mapped into quasi-circular orbits and that the new angle $\tilde \phi$ remains a rotation angle, which defines the secular average in the new coordinates.
Under this gauge transformation, the first term in Eq.~\eqref{eq:L} reads as
\begin{equation}
    \int_0^{2\pi}  \bar{L}|_{\gamma} d\tilde{\phi} \to \int_{0}^{2\pi} \bar{L}|_{\gamma}d\phi +\eta \int_0^{2\pi} \bar{L}|_{\gamma}d\chi = 2\pi \bar{L} ~,
\end{equation}
where we used the periodicity of $\chi$ and the fact that $\bar{L}$ does not depend on $\phi$.
The second term in Eq.~\eqref{eq:L}, under the gauge transformation in \eqref{eq:gauge}, transforms as 
\begin{equation}
    \int_0^{2\pi} L_1|_{\gamma} d\tilde{\phi} \to \int_0^{2\pi} L_1|_{\gamma} d\phi +\eta \int_0^{2\pi} L_1|_{\gamma}d\chi ~.
\end{equation}
The second integral in the expression above does not vanish in general, since $L_1$ depends on $\phi$. However, we can neglect it because the second integral will be multiplied by $\eta^2$ and therefore it is of higher order.
Putting the pieces together we have 
\begin{equation}
    \langle L \rangle \simeq \int_0^{2\pi}  \left(\bar{L}+\eta L_1\right)|_{\gamma} d\tilde{\phi} \to 2\pi\bar{L}+\eta \int_0^{2\pi}L_1|_{\gamma} d\phi  ~,
\end{equation}
thus $\langle L \rangle$ is gauge invariant under coordinate transformations of order $\mathcal{O}(\eta)$  which are $2\pi$-periodic in $\phi$. 
Along the same line of reasoning, one can prove the gauge invariance of $\langle u^{\phi} \rangle$ and $\langle u^t \rangle$. Since the orbital frequency for a quasi-circular orbit is defined by 
\begin{equation}
\Omega=\frac{u^{\phi}}{u^t}~,
\end{equation}
we conclude that $\langle \Omega \rangle$ is also gauge-invariant under coordinate transformations 
of order $\mathcal{O}(\eta)$ 
which are $2\pi$-periodic in $\phi$.

\subsection{Tidal effects around the ISCO}

The ISCO for massive test-particles is completely characterised by three parameters: its radius, energy and angular momentum. 
It is defined as an extreme point of the Hamiltonian \eqref{eq:averageH}, namely 
\begin{equation}
\label{eq: ISCO_cond}
\begin{split}
    \langle H\rangle \vert_{r=r_{\rm  ISCO}}=-\frac{1}{2}~,
    \\
    \frac{d\langle H\rangle}{dr}\bigg\vert_{r=r_{\rm ISCO}}=0~, 
    \\
    \frac{d^2 \langle H\rangle }{dr^2}\bigg\vert_{r=r_{\rm ISCO}}=0~.
\end{split}
\end{equation}
Using these conditions and keeping only terms proportional to $\eta$, it is possible to compute the secular effects caused by the tidal perturbations to the energy, angular momentum and radius of the Schwarzschild ISCO.

We assume that observables are expanded around their unperturbed values. Physically, this is equivalent to assume that tidal (secular) effects are all proportional to the tidal parameter $\eta$.~\footnote{We recall that we consider only up to first order contributions in the small-tide approximation.} This assumption also defines the numerical values of the tidal corrections. Tidal corrections to the radius,\footnote{Which is not a gauge-invariant quantity; see discussion at the end of this section.} the averaged energy and angular momentum read as~\footnote{From now on, we will drop the symbol of the secular average $\langle \cdot \rangle$ for the sake of presentation.}
\begin{equation}
\begin{split}
    r_{\rm ISCO}&\simeq r_0+\eta\, r_1 \,,
    \\
    E_{\rm ISCO}&\simeq E_0+\eta\, E_1\,,
    \\
    L_{\rm ISCO}&\simeq L_0+\eta\, L_1\,.
\end{split}
\end{equation}
By solving Eqs.~\eqref{eq: ISCO_cond} at  leading order one can determine the value of $(r_0,E_0,L_0)$, respectively the value for the radius, the energy and the angular momentum of the ISCO for an unperturbed Schwarzschild black hole. They are
\begin{equation}
\label{eq: isco0}
    r_0=6~ M~,
    ~~
    E_0=\frac{2\sqrt{2}}{3}~,
    ~~
    L_0=2\sqrt{3}~M~.
\end{equation}
At the first order in $\eta$, the first corrections to the ISCO quantities are given by
\begin{equation}
\label{eq: isco1}
    r_{1}=3072~ M~,
    ~~
    E_1=-\frac{152 \sqrt{2}}{3}~,
    ~~
    L_1=-348 \sqrt{3}~ M~.
\end{equation}
Note that we fixed our conventions for $\eta$ in order to precisely reproduce the same numerical values of $(r_1, E_1, L_1)$ previously obtained in Ref.~\cite{Yang:2017aht}. However, while the results of Ref.~\cite{Yang:2017aht} are only valid in the weak-field approximation where $\hat{r}\gg M_{\star}$ and on the equatorial plane $\hat \theta=\pi/2$, our results are more general and hold for any value of $\hat{r}$ and $\hat \theta$, as we discussed earlier in Sec.~\ref{sec: secular H}. 

It is also possible to compute the shift in the ISCO orbital frequency. In general, for quasi-circular orbits, the orbital frequency can be determined by means of the ratio \cite{Detweiler:2008ft,Yang:2017aht}
\begin{equation}
    \label{eq: Omega_def}
    \Omega^2=\left(\frac{u^{\phi}}{u^{t}}\right)^2=\frac{M}{r^3}-\frac{\left(r-3 M\right)}{2r^2}u^\mu u^\nu\partial_r \langle h_{\mu\nu}\rangle~,
\end{equation}
where $u^{\mu}$ are the components of the four-velocity \eqref{eq: p-epsilon}. To first order in $\eta$, we obtain
\begin{equation}
\label{eq: OmegaEXP}
    \Omega_{\rm ISCO}\simeq \Omega_0+ \eta~\Omega_1~,
\end{equation}
where~\footnote{Notice that this result agrees with  Ref.~\cite{Cardoso:2021qqu} (but not with Ref.~\cite{Yang:2017aht}),  after a rescaling of -1/2 of the $\eta$ parameter. For the ease of comparison, our radial configuration (see Fig.~\ref{fig:4configurations}) is called polar companion configuration in Ref.~\cite{Cardoso:2021qqu}: this can be obtained in the weak-field limit $\hat r \gg M_*$ and for $\beta=\pi/2$ and $\chi =- \pi/2$.}
\begin{equation}
    \label{eq: Omega_shift}
    M~\Omega_0=\frac{1}{6 \sqrt{6}},\quad M~\Omega_1=-\sqrt{\frac{2}{3}}\frac{491}{6}~.
\end{equation}
This gives the shift induced by the tidal fields in the orbital frequency of the ISCO. 

Following Ref.~\cite{Detweiler:2008ft}, the angular frequency $\Omega$ can be used to compute a gauge-independent measure of the radial separation between the Schwarzschild black hole and the test particle. One defines
\begin{equation}
    R_\Omega=\left(\frac{M}{\Omega^2}\right)^{1/3}~,
\end{equation}
so that according to Eqs.~\eqref{eq: OmegaEXP} and \eqref{eq: Omega_shift}
\begin{equation}
    R_\Omega\simeq \frac{2^{2/3}M}{\Omega^{2/3}_0}\left(1- \frac{2}{3}\eta\frac{\Omega_1}{\Omega_{0}}\right)= 6 M+ 3928\eta~ M~.
\end{equation}
We notice that this gives a different radial shift than in Eq.~\eqref{eq: isco1}. However, this is not surprising as the radial shift of Eq.~\eqref{eq: isco1}, unlike the above, is not gauge invariant.

\subsection{Tidal effects around the photon sphere}

The photon sphere around a Schwarzschild black hole is  composed by the last stable circular orbits for massless test-particles. Differently from the case of the ISCO, this orbit is only specified by two parameters: the photon sphere radius and the impact parameter $b=L/E$.
A previous analysis of the photon sphere properties in a tidal environment can be found in Ref.~\cite{Cardoso:2021qqu}, under more limited assumptions than the ones considered in this paper.
From the secular Hamiltonian \eqref{eq:averageH}, one enforces the conditions
\begin{equation}
\label{eq: ISCO_cond1}
\begin{split}
    \langle H\rangle \vert_{r=r_{\rm PS}}&=0~,
    \\
    \frac{d\langle H\rangle}{dr}\bigg\vert_{r=r_{\rm PS}}&=0~.
\end{split}
\end{equation}
By expanding the kinematic quantities in the tidal parameter $\eta$ to retain only the leading contribution of the tidal secular effects in the small-tide approximation, one obtains
\begin{equation}
\begin{split}
    r_{\rm PS}&\simeq r_0+\eta~ r_1~,
    \\
    b_{\rm PS}&\simeq b_0+\eta ~b_1~,
\end{split}
\end{equation}
where the unperturbed values for the Schwarzschild black hole are obtained by solving \eqref{eq: ISCO_cond1} at the leading order
\begin{equation}
\label{eq: LR_0}
    r_0=3~M~,
    ~~
    b_0=3\sqrt{3}~M~.
\end{equation}
Similarly, the tidal corrections are given by
\begin{equation}
\label{eq: LR_1}
    r_{1}=-30~M~,
    ~~
    b_{1}=30\sqrt{3}~M~.
\end{equation}
This results generalize the one obtained in Ref.~\cite{Cardoso:2021qqu} for the special configuration of polar companions (equivalent to our radial configuration), after a rescaling of $\eta$.

Again, the orbital frequency at the photon sphere at first order in the tidal corrections can be computed in general from
\begin{equation}
    \Omega=\frac{u^{\phi}}{u^{t}}=\frac{1}{b}~,
\end{equation}
which at  first order in $\eta$ yields to
\begin{equation}
    \Omega_{\rm PS}\simeq \Omega_0+ \eta~ \Omega_1~.
\end{equation}
By means of Eqs.~\eqref{eq: LR_0} and \eqref{eq: LR_1}, one directly obtains the shift in the frequency of the photon sphere, given by
\begin{equation}
    M~\Omega_0=\frac{1}{3 \sqrt{3}}~,~~ M~\Omega_1=-\frac{10}{3\sqrt{3}}~.
\end{equation}
%

\section{Conclusions and outlook}
\label{Concl}

We conclude by summarising our new results and discussing further developments.
In Sec.~\ref{sec: tidal construction}, we retraced the computation performed in Ref.~\cite{Marck:1973} for the construction of the Marck's tetrad, defining a local inertial frame which is parallel-transported around a time-like geodesic in Kerr spacetime. Tidal effects induced by a Kerr black hole are obtained by projecting the Weyl tensor on certain components of the Marck's tetrad. 
While the components of the rank-2 tensor $C_{ij}$ were computed in Marck's paper~\cite{Marck:1973}, the components of the rank-3 tensor $C_{ijk}$ were previously known only on the equatorial plane of a Kerr black hole~\cite{Alvi:1999cw,Poisson:2003nc}. This paper therefore fills the gap in the literature: the explicit expressions for $C_{ijk}$ are given in Eq.~\eqref{Cijk_Kerr}. Our result is valid for generic angles $\hat{\theta}$ and for arbitrary time-like geodesics in the Kerr spacetime.

In Sec.~\ref{Sec:triplete}, we found a natural application of the tidal tensors computed in the previous section in the modeling of a hierarchical three-body system in General Relativity. We considered a 3-body system describing a supermassive rotating black hole of mass $M_*$ and an EMR binary system, made of a non-rotating black hole of mass $M \ll M_*$ and a smaller companion of mass $m \ll M$, which gravitates around the supermassive black hole. In order to go beyond the post-Newtonian approximation, in which the three bodies are sufficiently distant from each other to be treated as point-like masses, and capture strong general relativistic effects, one can model the region around the non-rotating black hole in terms of a tidally-deformed Schwarzschild spacetime. To this aim, it is convenient to decompose the tidal tensor in terms of irreducible representations of the rotation group, so as to construct ``electric" $\mathcal{E}$ and ``magnetic" $\mathcal{B}$ quadrupole tidal moments, that encode the leading-order deformations to the Schwzarschild metric immersed in a generic tidal environment~\cite{Poisson:2009qj}.
By approximating the motion of the smallest body as that of a test-mass, it is possible to take into account all the possible configurations of the binary system by introducing two Euler's angles. Another new result obtained in this work is the explicit expressions for the electric and magnetic quadrupole tidal moments given in Eqs.~\eqref{general E}-\eqref{generalB}, that take into account arbitrary orientations of the binary system with respect to the source of the tidal deformations. We remark that these expressions are valid for arbitrary sources of tidal effects. This can be of interest for numerical simulations and analytical study of binary systems immersed in a tidal environment.
For the case of a supermassive Kerr black hole, the tidal moments \eqref{general E} and \eqref{generalB} together with our result in Sec.~\ref{sec: tidal construction} allow us to analytically compute tidal effects induced by a Kerr black hole in full generality. 
The hierarchy of masses makes it natural to study the dynamics of the binary system in the secular approximation. As first pointed out in Ref.~\cite{Yang:2017aht}, the tidal effects perturb the secular Hamiltonian for the binary system. Remarkably, at the quadrupole approximation, the tidal perturbation can be recast into an effective perturbative parameter $\eta$. The main result of Sec.~\ref{sec: secular H} is a general expression for $\eta$ given in Eq.~\eqref{etageneral}. It holds at the quadrupole order in the small-tide regime and in the secular approximation, and it models the deformed secular dynamics of a binary system. Our $\eta$ generalises results obtained in Ref.~\cite{Yang:2017aht} and Ref.~\cite{Cardoso:2021qqu} to arbitrary orientations of the binary system and tidal effects induced by a rotating black hole, including the strong gravity regime. 
Tidal deformations induce changes in certain gauge-invariant quantities characterising the EMR binary systems, such as the orbital frequency. Such tidal deformations induced by the environment are completely encoded in the effective perturbative parameter $\eta$. We devoted Sec.~\ref{sec: secular shift} to the study of such shifts in the case of marginally stable orbits for massive (ISCO shifts) and massless (photon sphere shifts) test-particles. We also addressed the issue of the gauge invariance of the shifts in the secular approximation.
While we focus on the case of a Kerr black hole as the perturber, one can also use our expressions with general tidal moments.
For a Kerr perturber, the expression for $\eta$ (see Eq.~\eqref{etaKerr}) shows the rich phenomenology of the triple system: it combines the parameters of the background Kerr metric ($M_*$ and $a$), the location of the geodesic where the binary system is located ($\hat r,~\hat \theta,~ K $), and the Euler angles that capture the geometric orientation of the binary system with respect to the Kerr perturber ($\beta$ and $\chi$). Our parameter $\eta$ includes strong general relativistic effects of an EMR binary system which is affected by the presence of a large Kerr black hole, and considerably generalises the setup considered in Refs.~\cite{Yang:2017aht} and~\cite{Cardoso:2021qqu} beyond the weak-field regime and for arbitrary configurations. As an example of a regime which was previously overlooked in the literature about tidally deformed binaries, in Sec.~\ref{Sec:equatorial}, we focused on the case in which the EMR system is placed on the ISCO of the Kerr background. We also derived configurations of the EMR system for which the tidal effects vanish in the secular approximation, generalising the findings of Ref.~\cite{Yang:2017aht}.
There is a number of directions in which this work can be further extended, and for which the results obtained here can be of interest. 
In this paper, we analyze triple systems whose dynamics is stationary in time and restricted to circular orbits. This implies that we do not have gravitational waves in our setup.
We also work in the leading quadrupole approximation for the tidal effects.
The setup in this paper, though simplified, is useful to get analytic results and it should be considered as a first step towards a more realistic scenario that can be relevant for astrophysical interest.
An extension of this work would include higher-order effects beyond the quadrupole approximation~\cite{Will:2020tri} and the stationary regime. It would be interesting to further develop waveforms from triple hierarchical systems~\cite{PhysRevD.101.104053,Bonetti:2017hnb} and approaches to effective description thereof~\cite{Kuntz:2021ohi,Kuntz:2022onu}.
Another natural development would be extending this study to the case in which the primary companion of the EMR is a Kerr black hole. The metric for a rotating black hole deformed by tidal effects has been derived in full generality in Ref.~\cite{Yunes:2005ve} by solving the Teukolsky equation and using metric reconstruction techniques. Due to the very complicated structure of that metric, a simplified version obtained in the small-spin regime has been obtained in Ref.~\cite{Poisson:2014gka}, explicitly written in terms of tidal quadrupole moments. This is sufficient to capture all the main important features of spacetimes with non-vanishing angular momentum, and can lead to an even richer phenomenology -- including couplings between the spins of the two black holes -- possibly already at the level of the secular dynamics.
A third interesting direction concerns the analysis of eccentric binary systems subject to tidal deformations. For this specific case it is probably more convenient to use the action-angle variables formalism \cite{Schmidt:2002qk, Drasco:2005kz, Glampedakis:2005cf,Hinderer:2008dm}. This would allow us not only to extend our computation to the case of elliptic orbits for the test particle in the binary system, but also to study the precession of the orbits around the Schwarzschild black hole and the presence of possible resonances in the binary system \cite{Naoz:2012bx,Brink:2015roa}.

\section*{Acknowledgments}

We thank P. S. Cole, B. Liu and J. Samsing for interesting discussions.
We thank V. Cardoso for useful comments on the manuscript.
G.G.~and M.O.~acknowledge support from Fondo Ricerca di Base 2020 (MOSAICO) and 2021 (MEGA) of the University of Perugia.
The work of T.H. is supported in part by the project “Towards a deeper understanding of black holes with non-relativistic holography” of the Independent Research Fund Denmark (grant number DFF-6108-00340). 
The work of R.O.~is supported by the R\'egion \^Ile-de-France within the DIM ACAV$^{+}$ project SYMONGRAV (Sym\'etries asymptotiques et ondes gravitationnelles). 
G.G. and R.O.~thank the Niels Bohr Institute for hospitality at different stages of this project. T.H. thanks University of Perugia for hospitality.


\bibliographystyle{apsrev4-1}
\bibliography{gwbib}

\end{document}